\documentclass[twocolumn]{aastex63}
\usepackage{graphicx}
\usepackage{graphics}
\usepackage{amssymb}
\usepackage{bm}
\usepackage{times}
\usepackage[flushleft]{threeparttable}
\usepackage{booktabs}
\usepackage{footnote}
\usepackage{multirow}

\usepackage{lineno}

\definecolor{green}{rgb}{0.0, 0.6, 0.6}

\newcommand{\hi}{H{\sc i}}
\newcommand{\zh}{$\rm Z_{HI}$}
\newcommand{\zs}{$\rm Z_{SFR}$}
\newcommand{\zsm}{$\rm Z_{SFR,med}$}

\shorttitle{Metal enrichment of ionised and atomic gas}
\shortauthors{Arabsalmani et al.}
\begin{document}


\title{A comprehensive study on the relation between the metal enrichment of ionised and  atomic gas  in star-forming galaxies}

\correspondingauthor{Maryam Arabsalmani, Vera-Rubin Fellow}
\email{maryam.arabsalmani@origins-cluster.de}

\author{M. Arabsalmani} 
\affiliation{Excellence Cluster ORIGINS, Boltzmannstra{\ss}e 2, 85748 Garching, Germany}
\affiliation{Ludwig-Maximilians-Universit\"at, Schellingstra{\ss}e 4, 80799 M\"unchen, Germany}

\author{L. Garratt-Smithson} 
\affiliation{International Centre for Radio Astronomy Research (ICRAR), University of Western Australia, 35 Stirling Hwy, Crawley, WA 6009, Australia}
\affiliation{ARC Centre of Excellence for All Sky Astrophysics in 3 Dimensions (ASTRO 3D)}

\author{N. Wijers} 
\affiliation{Leiden Observatory, Leiden University, Niels Bohrweg 2, NL-2333 CA Leiden, The Netherlands}
\affiliation{CIERA and Department of Physics and Astronomy, Northwestern University, 1800 Sherman Ave, Evanston, IL 60201, USA}

\author{J. Schaye} 
\affiliation{Leiden Observatory, Leiden University, Niels Bohrweg 2, NL-2333 CA Leiden, The Netherlands}

\author{A. Burkert}
\affiliation{University Observatory Munich (USM), Scheinerstra{\ss}e 1, 81679 M\"unchen, Germany}
\affiliation{Max-Planck-Institut f\"ur extraterrestrische Physik (MPE), Giessenbachstr. 1, 85748 Garching, Germany}

\author{C.D.P. Lagos} 
\affiliation{International Centre for Radio Astronomy Research (ICRAR), University of Western Australia, 35 Stirling Hwy, Crawley, WA 6009, Australia}
\affiliation{ARC Centre of Excellence for All Sky Astrophysics in 3 Dimensions (ASTRO 3D)}
\affiliation{Cosmic Dawn Center (DAWN)}

\author{E. Le Floc'h}
\affiliation{CEA, IRFU, DAp, AIM, Universit\'e Paris-Saclay, Universit\'e Paris Cit\'e, Sorbonne Paris Cit\'e, CNRS, 91191 Gif-sur-Yvette, France}

\author{D. Obreschkow}
\affiliation{International Centre for Radio Astronomy Research (ICRAR), University of Western Australia, 35 Stirling Hwy, Crawley, WA 6009, Australia}

\author{C. Peroux}
\affiliation{European Southern Observatory, Karl-Schwarzschildstrasse 2, D-85748 Garching bei M{\"u}nchen, Germany}
\affiliation{Aix Marseille Universit\'e, CNRS, LAM (Laboratoire d'Astrophysique de Marseille) UMR 7326, 13388, Marseille, France}

\author{B. Schneider}
\affiliation{CEA, IRFU, DAp, AIM, Universit\'e Paris-Saclay, Universit\'e Paris Cit\'e, Sorbonne Paris Cit\'e, CNRS, 91191 Gif-sur-Yvette, France}


\begin{abstract}
We study the relation between the metallicities of ionised and atomic gas in star-forming galaxies at {$z=0-3$} using the EAGLE cosmological, hydrodynamical simulations. This is done by constructing a dense grid of sightlines through the simulated galaxies and obtaining the star formation rate- and H{\sc i} column density-weighted metallicities, $\rm Z_{SFR}$ and $\rm Z_{HI}$, for each sightline as proxies for the metallicities of ionised and atomic gas, respectively. We find $\rm Z_{SFR} \gtrsim Z_{HI}$ for almost all sightlines, with their difference generally increasing with decreasing metallicity. The stellar masses of galaxies do not have a significant effect on this trend, but the positions of the sightlines with respect to the galaxy centres play an important role: the difference between the two metallicities decreases when moving towards the galaxy centres, and saturates to a minimum value in the central regions of galaxies, irrespective of redshift and stellar mass. This implies that the mixing of the two gas phases  is most efficient in the central regions of galaxies where sightlines generally have high column densities of H{\sc i}. However, a high H{\sc i} column density alone does not guarantee a small difference between the two metallicities. 
In galaxy outskirts,  the inefficiency of the mixing of star-forming gas with \hi\  seems to dominate over the  dilution of heavy elements  in \hi\   through mixing with the pristine gas. 
We find good agreement between the available observational data and the $\rm Z_{SFR}$-$\rm Z_{HI}$ relation predicted by the EAGLE simulations. Though{, observed} regions with a {nuclear} starburst mode of star formation appear not to follow the same relation.   
\end{abstract}




\section{Introduction} 
\label{sec:int}

The content  of heavy elements in galaxies over cosmic time provides essential information on how galaxies form and evolve. The metal enrichment of galaxies is not only linked to their star formation history, but is also related to outflowing and infalling gas in galaxies, through which metals are lost or diluted with pristine gas \citep[e.g.,][]{Matteucci12-2012ceg..book.....M, Silk12-2012RAA....12..917S, Naab17-2017ARA&A..55...59N, Dayal18-2018PhR...780....1D}. Metallicity  is therefore one of the key properties of galaxies, and its relation with other galaxy properties provides crucial information on the mechanisms involved in galaxy formation and evolution \citep[e.g.,][]{Tremonti04-2004ApJ...613..898T, Mannucci09-2009MNRAS.398.1915M, Mannucci10-2010MNRAS.408.2115M, Bothwell13-2013MNRAS.433.1425B, Moller13-2013MNRAS.430.2680M, Zahid14-2014ApJ...791..130Z, Yabe15-2015PASJ...67..102Y, Arabsalmani15-2015MNRAS.446..990A, Arabsalmani18-2018MNRAS.473.3312A}.

The metallicity of a galaxy is defined as the ratio between the abundance of the elements heavier than helium (metals) and that of hydrogen$+$helium in the galaxy. This is  observationally obtained from the abundances of elements in the  stellar populations or in the  interstellar medium (ISM) of a galaxy. The latter measurements are derived  separately for the ionised and the atomic   gas phases. The ionised gas metallicities are generally measured from the ratios of strong emission lines in the spectra of  galaxies. Such measurements are based on calibrations which are only well-constrained at low redshifts (and only within a metallicity range) and may not hold at higher redshifts { \citep[e.g., see][]{Maiolino08-2008A&A...488..463M, Steidel14-2014ApJ...795..165S}}. Moreover,  it is shown that different calibrations give rise to metallicities that differ significantly for a given galaxy, even at low redshifts { \citep[][]{Kewley08-2008ApJ...681.1183K}}. 
The atomic  gas metallicity measurements are instead based on the ratio between the column densities of heavy elements and atomic  hydrogen derived from ISM absorption lines against background continuum sources.   These measurements provide  more reliable estimates of metal enrichment up to very high redshifts  \citep[$z > 5$,][]{Sparre14-2014ApJ...785..150S, Hartoog15-2015A&A...580A.139H}, and  do not rely on any calibration (negligible ionisation corrections and also dust corrections are assumed for these measurements). However, they provide the metallicity measurement only in a narrow beam along the bright background source -- typically the quasars (QSOs), or the afterglows of Gamma Ray Bursts (GRBs), unlike the  case of ionised gas metallicities measured from emission-lines where  the metallicity measured is representative of an entire galactic region.  
{Note that these ratios measure the metal enrichment of the  atomic gas, and not that of the combined neutral molecular and atomic gas.}

Despite the crucial role of the metal enrichment of galaxies in our understanding of galaxy formation and evolution, measurements of  the true metallicities  of galaxies remain  uncertain.  
It is not clear  whether the metallicities derived from the stellar populations are / should be   the same  as those obtained for the gas. 
Even the metallicities derived for different gas phases (ionised and atomic) in a galaxy  might yield different values. Understanding how these measurements relate to each other will not only provide essential information on the mixing of metals in different gas phases, but is also required for interpreting the measurements  derived  with  different methods. 

To date, there are only a handful of measurements for the metallicities of both ionised and atomic phases of gas in the same galaxies. These are mainly obtained for intervening galaxies along quasar sightlines, showing  deviations between the two measurements \citep[see e.g.,][]{Christensen14-2014MNRAS.445..225C, Friis15-2015MNRAS.451..167F, Peroux16-2016MNRAS.457..903P, Rahmani16-2016MNRAS.463..980R, Hamanowicz20-2020MNRAS.492.2347H}. {In several cases the measurement are obtained for nearby galaxies \citep[e.g.,][]{Heckman01-2001ApJ...554.1021H, Thuan02-2002ApJ...565..941T, Thuan05-2005ApJ...621..269T, Cannon05-2005ApJ...618..247C, Lebouteiller-2009A&A...494..915L, Lebouteiller13-2013A&A...553A..16L, Hernandez12-2021ApJ...908..226H}.}
Given the rarity of these measurements, cosmological hydrodynamical simulations \citep[e.g,][]{Schaye2015, Pillepich18-2018MNRAS.473.4077P, Maio22-2022A&A...657A..47M} offer a natural test-bed to explore the relationship between the metallicities of ionised and atomic gas which can also be confronted to the few observational constraints. 

In this paper we present a detailed and comprehensive analysis of the relation between the atomic and ionised gas metallicities in star-forming galaxies in the redshift range $0\leq z \leq 3$ 
and investigate the  dependence of this relation on various parameters such as the global properties of galaxies as well as  the local properties of regions of interest. For this study we use   the EAGLE (Evolution and Assembly of GaLaxies and their Environments) set of cosmological hydrodynamical simulations \citep{Schaye2015}. We describe the numerical methods  in Section \ref{sec:sim}. The results and discussion   are presented in Section \ref{sec:res}. We  summarise our findings in Section \ref{sec:sum}. 
{We use a solar metallicity value of 0.0127 throughout this work. This is the default value used in \textsc{cloudy}~v7.02 (last documented in \cite{cloudy}), with element abundances from \cite{grevesse_sauval_1998}, \cite{allendeprieto_lambert_asplund_2001}, \cite{holweger_2001}, and \cite{allendeprieto_lambert_asplund_2002}.}



\section{Numerical Methods}
\label{sec:sim}

\subsection{Simulations}

This paper uses the publicly available cosmological hydrodynamical simulation EAGLE RecalL025N0752 \citep{Schaye2015, Crain15-2015MNRAS.450.1937C,  McAlpine16-2016A&C....15...72M, eagle-team_2017}. 
The correlations between metallicity and other galaxy properties -- such as stellar mass, star formation rate (SFR), specific star formation rate, gas fraction -- have been extensively studied using EAGLE \citep[e.g.,][]{Lagos16-2016MNRAS.459.2632L, DeRossi2017, Trayford2019, Vanloon21-2021MNRAS.504.4817V}, with authors finding a good agreement with observed trends down to stellar masses of 10$^9$ M$_{\odot}$ using the EAGLE run RecalL0025N0752. This is the highest resolution EAGLE simulation and is consequently the simulation we use for this analysis. Particularly relevant to this work, EAGLE has been shown to reproduce both the observed radial metallicity gradients within the central part of galaxies \citep[$<$ 0.5 times the half-mass radius;][]{Collacchioni2020}, along with $z \sim 0$ metallicity gradients within star-forming regions of disc galaxies \citep[within observed scatter;][]{Tissera2019}, and the redshift evolution of the metallicity gradient within star-forming regions \citep[although deviations in scatter were seen between the simulations and observations;][]{Tissera2022}. However, it is worth noting the recent paper by \citet{Yates2021}, where it was highlighted that the cosmic metal density of the neutral gas phase at high redshift ($3 \lesssim z \lesssim 5$) is significantly lower in existing cosmological simulations (including EAGLE), than observations of Damped Lyman-$\alpha$ systems (systems with N(\hi) $\geq 2 \times 10^{20}$ cm$^{-2}$) imply. This tension was seen across different cosmological simulations, hence it is not clear whether one simulation is favoured over another for this type of metallicity study (note that the study  presented in this paper is focused on $z\leq$ 3).  
Our choice to use EAGLE for this analysis was further motivated by the fact the RecalL025N0752 run has been shown to reproduce the $z$~$= 0$ global H$_2$ scaling relations \citep[e.g.,][]{Lagos2015}, alongside the \hi\ scaling relations \citep[e.g.,][]{Crain2017} and the \hi\ column density distribution function \citep[CDDF; e.g.,][]{Rahmati2015}.

\begin{figure*}
\centering
\includegraphics[width= 0.9\textwidth, angle = 0, trim={0 0 0 0}]{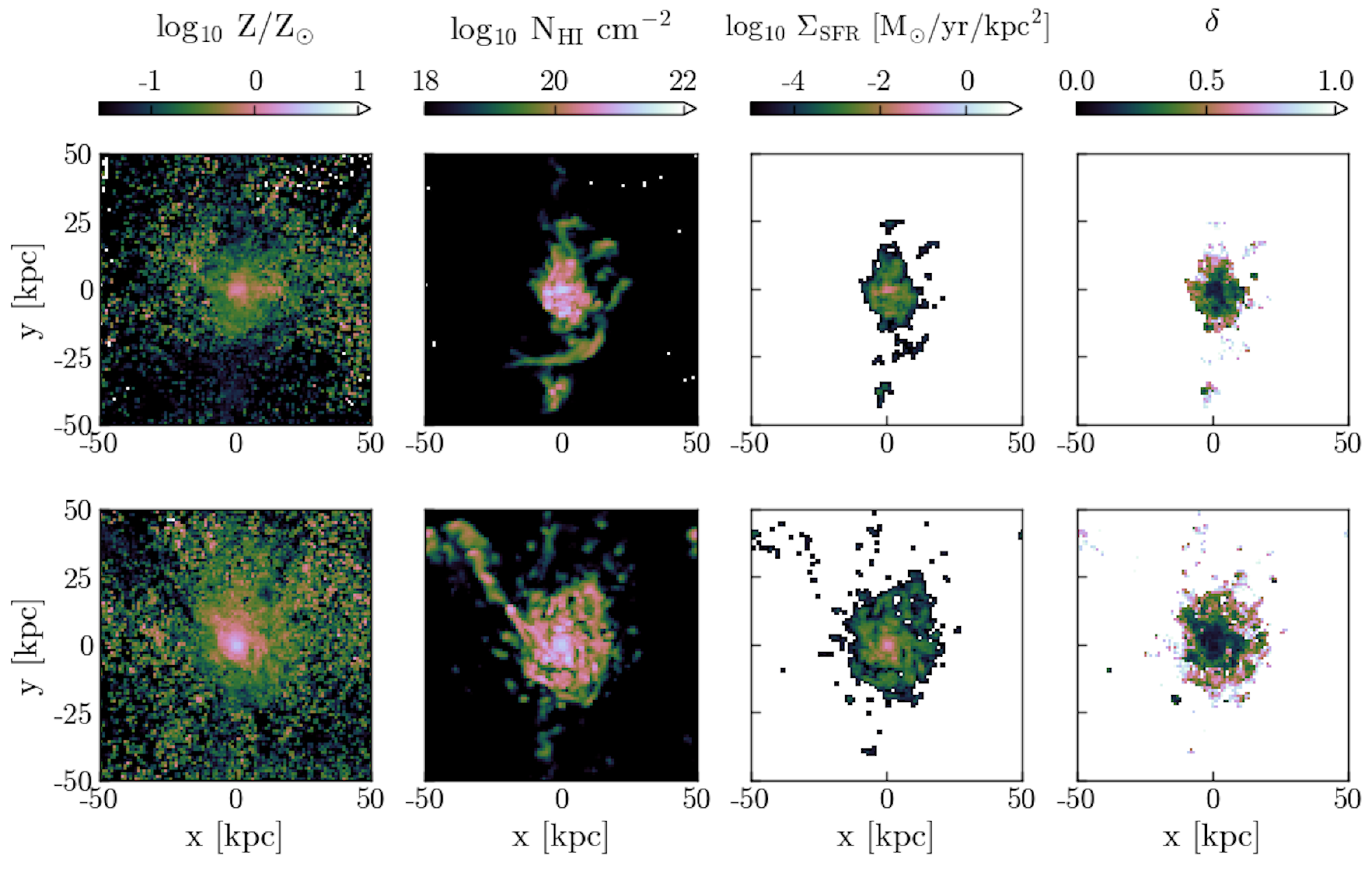}
\caption{  
Example of the construction of artificial sightlines about two galaxies of stellar mass 2$\times 10^9$ M$_{\odot}$ (top row) and 9$\times 10^{9}$ M$_{\odot}$ (bottom row). 
The four columns from left to right show the maps of  gas   metallicity, the \hi\ column density, the star formation rate surface density,  and $\rm log_{10}(Z_{SFR,med}/Z_{HI})$ defined as $\delta$ (see Section \ref{sec:res}), respectively. Each grid has a set pixel size of 2 kpc. The side length is determined by 2R$_{200}$, though in the interest of comparison, only the inner 100 kpc for each galaxy is shown here. 
\label{fig:maps}}
\end{figure*}

The simulations follow the evolution of a cubic volume of side-length  25 cMpc  of the universe to $z$~$= 0$, where  c  notes `comoving'. 
{They use a dark matter particle mass  of $1.21 \times 10^6 \, \rm{M}_{\odot}$ and an (initial) gas particle mass  of $2.26 \times 10^5 \, \rm{M}_{\odot}$.}
These simulations   use an updated version of the hybrid N-body Tree-PM/smooth particle hydrodynamics (SPH) code {\sc GADGET 3}, presented in \citet{Springel2005}. The modifications to the numerical methods in {\sc GADGET} -- i.e. the inclusion of an artificial thermal conduction switch \citep{Price2008}, a viscosity switch \citep{Cullen2010}, a time step limiter \citep{Durier2012} and the pressure-entropy formalism described in \citet{Hopkins2013} -- are collectively known as ``Anarchy" and are described in the appendix of \citet{Schaye2015} and in \citet{Schaller15-2015MNRAS.454.2277S}. The EAGLE simulations also model processes such as star formation and stellar/AGN feedback using sub-grid prescriptions, where the feedback efficiencies are calibrated \citep[][]{eagle_calibration}  to match the $z$~$= 0$ galaxy stellar mass function (GSMF), the sizes of disk galaxies, and the observed black hole mass -- stellar mass relation.

The star formation and metal enrichment/stellar-mass-loss sub-grid prescriptions are of particular importance to this work. Within the simulation, stellar populations are represented by star particles. These are formed using a metallicity-dependent density threshold designed to track the transition from the warm, atomic to the cold, molecular interstellar gas-phase \citep[first described in][]{Schaye2004}, alongside a pressure-dependent star formation rate that reproduces the observed Kennicutt-Schmidt star formation law  \citep[as described in][]{Schaye2008}. Star particle formation is stochastic, with a probability linked to the star formation rate of a gas particle. 
Further, as described in \citet{Wiersma09-2009MNRAS.399..574W}, the stellar-mass-loss for each star particle is approximated using the metallicity-dependent lifetimes presented in \citet{Portinari1998} and the AGN/stellar wind/core-collapse supernovae yields given in \citet{Marigo2001} and \citet{Portinari1998}, on top of the mass-loss associated with Type 1a supernovae. EAGLE tracks 9 elements individually; H, He, C, N, O, Ne, Mg, Si and Fe \citep[][]{wiersma_etal_2009_insim}. These elements are distributed across the neighbours of each star particle using the SPH kernel weighting. 
For this paper we use the metallicities associated with each particle, as opposed to the SPH-smoothed version of the metallicities (those used in the cooling calculations of the simulation, though not when calculating the star formation rate of each particle \citep[see][]{EAGLE17-2017arXiv170609899T}. 
{This is because in eq. 2 we  are interpolating to a grid, and hence  by smoothing the metallicities we would smooth twice wich in return  would artificially decrease the resolution. } 
Further, when we do use the SPH-smoothed version of the particle metallicities we see an artificial reduction in scatter at high impact parameters/low densities, driven by the fact the metallicity is smoothed over larger distances in the outskirts of the galaxy/CGM. These differences highlight the constraints of the simulation - since the resolution of SPH simulations naturally follows mass, the CGM/IGM remain poorly resolved, hence the metallicity values of any \hi\ in these regions should be interpreted with caution. 
{For {completeness},  we reproduce all the plots presented in Section \ref{sec:res} using the SPH-smoothed version of the particle metallicities and present them in the Appendix.}

\subsection{The atomic gas phase}
Similar to other cosmological-scale simulations \citep[e.g.,][]{Dubois14-2014MNRAS.444.1453D, Pillepich18-2018MNRAS.473.4077P}, EAGLE does not have the resolution to follow the cold gas phase of the galaxy; instead it imposes a density-dependent pressure floor, normalised to 8000 K at n$_{\rm{H}} = 10^{-1}$ cm$^{-3}$. Therefore, the simulation needs to be post-processed in order to estimate the {neutral (atomic and molecular)} gas fraction of each gas particle. We do this using the method described in Figure 1 of \citet{GarrattSmithson2021}, based on previous works \citep[e.g.,][]{Lagos2015, Bahe2016, Crain2017, Diemer2018}. 
{Note that we obtain both the  atomic gas fraction and  molecular gas fraction separately and use the former in order to estimate the metallicity of the  atomic gas.} 
We summarise the method in the following five steps.
\\
(i) 
{First, we calculate the total non-ionized hydrogen fraction, which combines molecular hydrogen and \hi. We do this according to the prescription of \citet{Rahmati2013}, assuming a \citet{Haardt2012} photo-ionizing background. In this prescription, the photo-ionization rate is estimated using a density-dependent fitting function. The fitting function was calibrated using cosmological simulations that included radiative transfer calculations. This photo-ionization rate is then used to estimate the fraction of non-ionized hydrogen in each gas particle. In the next steps, we calculate the molecular hydrogen fraction in the non-ionized gas, which allows us to calculate the \hi\ fraction.}
\\
(ii) From here, the molecular-dissociating UV-flux is estimated for each gas particle. The method we utilise here depends on whether the gas particle is star-forming or not. If it is, the flux is estimated using the ratio between the SFR density of the particle and that of the solar neighbourhood \citep[as in][]{Lagos2015}. 
If it is non-star-forming, the particle uses a minimum value set by the \citet[][]{Haardt2012} UV background. Additionally, following the work by \citet[][]{Diemer2018},  later adapted for EAGLE in \citet[][]{GarrattSmithson2021}, we assume 10 per cent of the UV photons emitted by star-forming particles escape and propagate through the surrounding medium (assumed to be optically thin). This provides an additional source of UV for non-star-forming particles.
\\
(iii) We then calculate two theoretically-motivated density floors for the cold neutral medium; the first assumes a 2-phase ISM model \citep[following][]{Wolfire2003}. The second is the minimum density of the cold neutral medium required to maintain pressure equilibrium \citep[as described in][]{Krumholz2013}.
\\
(iv) Next, we take the maximum of these density floors and use this value, alongside the molecular-dissociating radiation field, to estimate the optical depth of dust grains \citep[following][]{Krumholz2009,Krumholz2013} -- fundamental to the formation of molecular hydrogen.
\\
(v) Lastly, we calculate the fraction of molecular hydrogen using both the UV-flux and the optical depth of dust, utilising the method described in \citet{McKee2010}. The atomic  gas, \hi,  defined in this study  excludes the molecular hydrogen   as well as helium.

\begin{figure}
\centering
\includegraphics[width= 0.5 \textwidth]{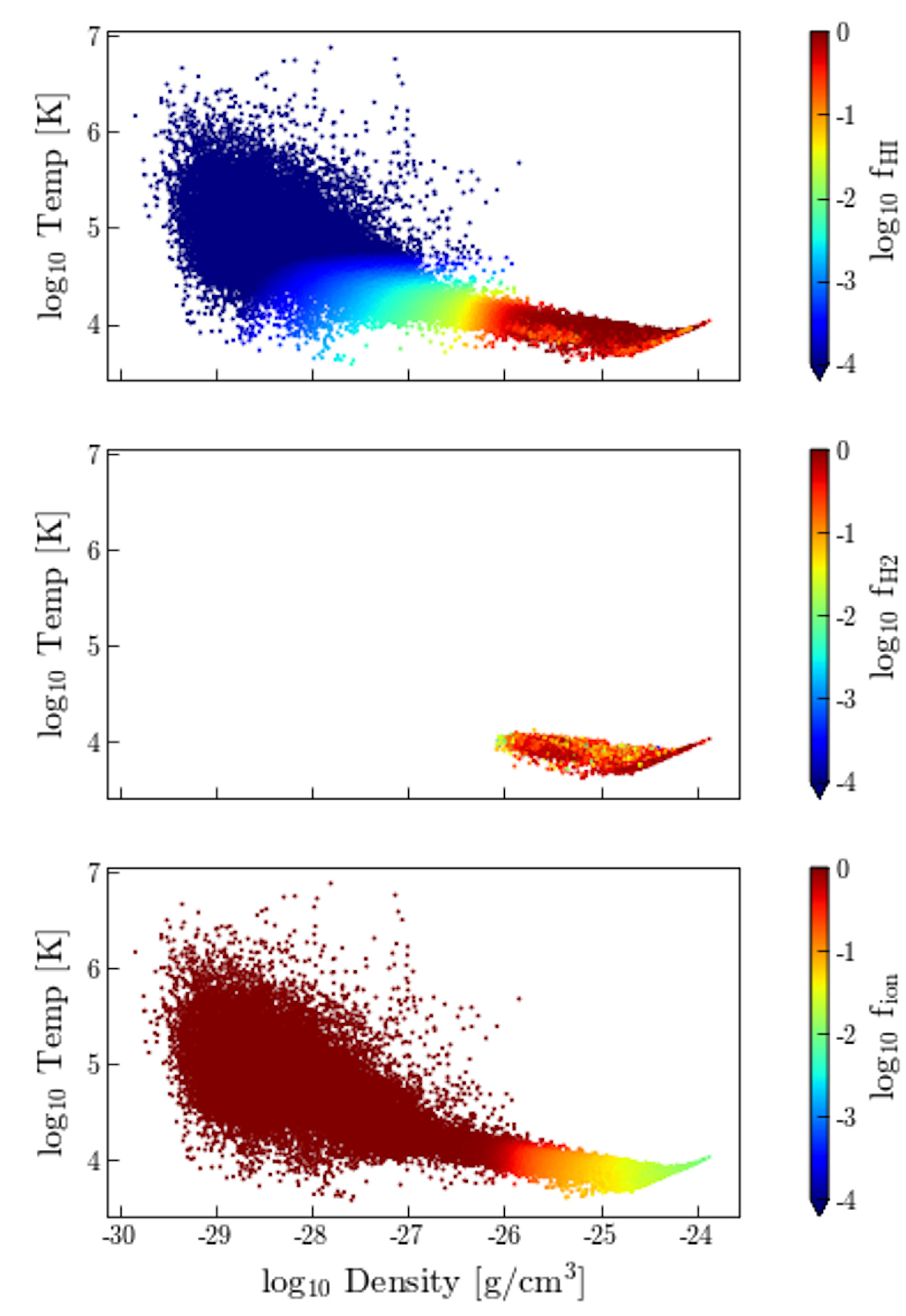}
\caption{ 
Gas phase diagram for the gas particles associated with 10 z$ = 0$ systems in our sample, with points colored by the HI mass fraction (f$_{\rm{HI}}$; upper plot), the molecular hydrogen mass fraction (f$_{\rm{H2}}$; middle plot) and the ionised gas mass fraction (f$_{\rm{ion}}$; lower plot) respectively. Only gas particles at radii below the virial radius of each system were included. {Note the  large molecular hydrogen mass fractions as well as the large  \hi\ mass fractions  at T $\sim 10^4$ K.} 
\label{fig:phase_plot}}
\end{figure}

{To illustrate our post-processing method, Figure \ref{fig:phase_plot} plots the gas phase diagram for all gas particles associated with 10 $z=0$ systems taken from our sample. Here we use all gas particles found {within} the virial radius  of each system. This will therefore include the circumgalactic medium alongside the interstellar medium. The top plot is coloured by the \hi\ mass fraction (f$_{\rm{HI}}$), the middle plot shows the molecular hydrogen mass fraction (f$_{\rm{H2}}$), while the lower plot uses the ionised gas fraction (f$_{\rm{ion}}$). Here we can see high f$_{\rm{HI}}$ and f$_{\rm{H2}}$ values are associated with the low temperature {(in particular, note the large values of  f$_{\rm{HI}}$ and f$_{\rm{H2}}$ at T $\sim 10^4$ K)}, high density regime, while the opposite is true for gas particles with {high} f$_{\rm{ion}}$. In particular, galaxies sitting on the artificial pressure floor (seen in the bottom right of the phase diagram), are typically H$_{2}$ dominated.}

\subsection{Constructing artificial sightlines}
\label{sec:s}
We constructed artificial sightlines for the simulation box using an N $=$ 8000$^2$ grid, with each grid cell representing a single sightline (each cell has a side length 3.125 ckpc). The box was then sliced into 5 in order to approximate typical velocity cuts used in observations \citep[as in][]{Rahmati2015}. 
{The metallicity of each pixel in our grid was calculated using a cloud-in-cell (CIC) algorithm. 
The CIC algorithm is a multi-linear interpolation scheme to evaluate fields on a regular grid based on irregularly placed
particles. 
Here each particle contributes a metallicity weighted by the HI mass or SFR that gas particle contributed to the cell, i.e.
\begin{equation}
    Z_{\rm{SFR}}= \frac{\sum_{i=1}^{N} \rm{SFR}_{\rm{cic},i} \times Z_{i}}{\sum_{i=1}^{N} \rm{SFR}_{\rm{cic},i}},
\end{equation}
where SFR$_{\rm{cic},i}$ is the cloud-in-cell contribution of the $i$th particle to the SFR of the cell and Z$_i$ is the metallicity of the $i$th particle. Similarly,
\begin{equation}
    Z_{\rm{HI}}= \frac{\sum_{i=1}^{N} \rm{MHI}_{\rm{cic},i} \times Z_{i}}{\sum_{i=1}^{N} \rm{MHI}_{\rm{cic},i}},
\end{equation}
where MHI$_{\rm{cic},i}$ is the cloud-in-cell contribution of the $i$th particle to the total HI mass of the cell.
The metallicities weighted by HI mass and star formation rate provide proxies for the metallicities of  atomic gas  and ionised gas, respectively, thus  we can compare the typical metallicity as measured using global cold gas in galaxies, with the metallicity of the star-forming regions.} 
{We mention again that throughout this paper, we use the metallicity of  atomic gas, as the neutral gas metallicities obtained from observations in the literature refer to the metal enrichment of only atomic gas.}  
{Given that the EAGLE simulations reproduce the Kennicutt-Schmidt star formation law  \citep[as described in][]{Schaye2008}, we expect the H$_2$-weighted metallicities to be very similar to  SFR-weighted metallicities. We investigate this and we indeed find this to be the case.}

In order to investigate the relation  between the  \hi-weighted metallicity and the  SFR-weighted metallicity on a galaxy-by-galaxy level, we also took all central galaxies with a stellar mass of $> 10^8$ M$_{\odot}$ out of RecalL025N0752 and created a grid centered on each individual galaxy, using a side-length of 2R$_{200}$ -- the physical radius within which density is 200 times the critical density of the Universe, calculated for each halo identified via the friends-of-friends algorithm.  Here we negate any projection-effects by explicitly rotating each system so that we are observing it perpendicularly to the angular momentum axis of the star particles. Further, the size of each pixel is set to 2 pkpc, independent of the value of R$_{200}$. This means there will be a smaller number of pixels for lower mass galaxies. As above, we then use this grid to interpolate $Z_{\rm{HI}}$, $Z_{\rm{SFR}}$, along with the \hi\ column density N(\hi) for each pixel. 
In observations, the impact parameter of a sightline, $b$,  is defined    as    the projected distance between the sightline and the  centre of its corresponding  galaxy.  
We too calculate the impact parameter, $b$, using the centre of each pixel in the 2D $x-y$ plane (where the $z$ direction is the rotation axis). 
Figure \ref{fig:maps} shows an example of this process for two galaxies of stellar mass 2$\times 10^9$ M$_{\odot}$ (top row) and 9$\times 10^{9}$ M$_{\odot}$ (bottom row). 

{In the  following section we present and discuss  our findings using the sightlines constructed for individual galaxies. This is to ensure that there is no more than one galaxy  per sightline, which is likely not  the case for sightlines in  the entire box (the sightlines constructed in the full box might correspond  to more than one galaxy). We note that the results show the  same trends no matter if we use all sightlines in the simulation box, or the sightlines produced using galaxies cut-out from the simulation.}



\section{results}
\label{sec:res}

\subsection{\zh--\zs\ plane of EAGLE galaxies}
\begin{figure}
\centering
\includegraphics[width=0.5 \textwidth]{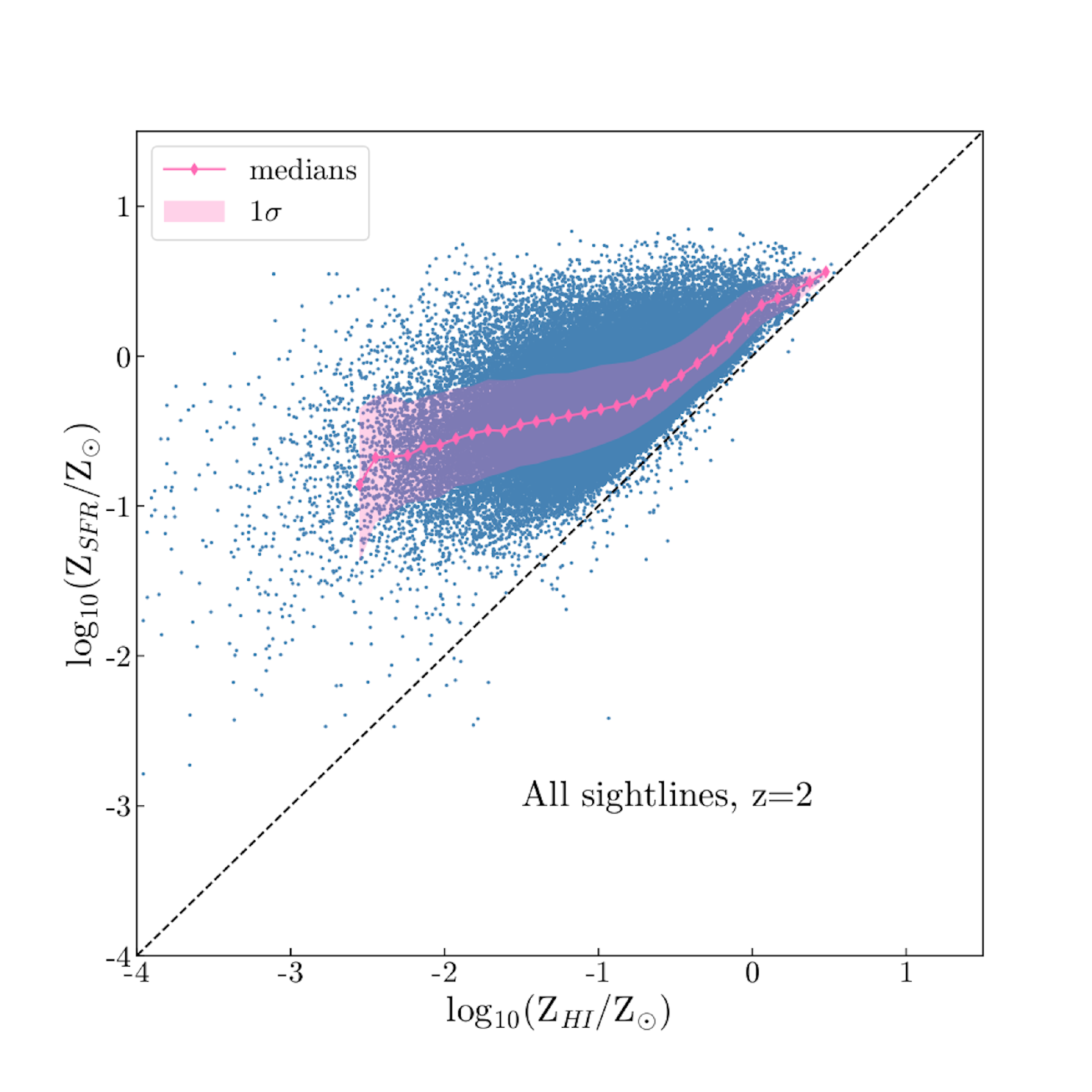}
\caption{ 
The \zh--\zs\ plane for all the sightlines in EAGLE galaxies at $z=2$. The medians of the \zs\ values in each \zh\ bin are shown with pink diamonds. The shaded area shows the  $1\sigma$ spread around the medians of the \zs\ values. The x=y line in the plane is presented by a dashed line.    
\label{fig:all-z2}}
\end{figure}

The SFR- and \hi- weighted metallicities (\zs\ and \zh\, respectively) of all the sightlines, excluding those  with zero SFR or  zero \hi,  in EAGLE galaxies at $z=2$ are presented in Figure \ref{fig:all-z2}. 
The first notable finding is that for almost all the sightlines, the \zs\ values are larger than the \zh\ values. 
This implies that the star-forming gas in all regions of the galaxies is more enriched with metals than the \hi\ gas located along the sightlines passing the same regions. 
It is also apparent that the  majority of the sightlines have \zs\ above $\sim$ 0.1  solar metallicity. For  the same sightlines the \zh\ values can be   an order of magnitude lower. 
The saturation of the \zs\ values could be a result of the local star formation enriching  the surrounding gas  to \zs\ $\gtrsim$ 0.1 $\rm Z_{\odot}$.

To undertake a more quantitative investigation, we use the medians of the \zs\ values, \zsm, within bins of \zh. The medians are marked  with the pink diamonds in Figure \ref{fig:all-z2}, and the shaded area in the plot shows the $1\sigma$ region around the median values. 
In each \zh\ bin, we measure the difference between   $\rm log_{10}(Z_{SFR,med}/Z_{\odot})$   and the central  $\rm log_{10}(Z_{HI}/Z_{\odot})$ value of  the bin, and refer to it as $\delta$. 
$\delta$  clearly increases  with a decrease in \zh\ and consequently a decrease in \zsm:  for a \zsm\ of  $\rm Z_{\odot}$, $\delta$ is about 0.2 dex, while its value increases to 1.2 dex for a \zsm\ of 0.3 $\rm Z_{\odot}$. Equivalently, $\delta$ increases from 0.2 dex to 1.2 dex when \zh\ decreases from  0.5 $\rm Z_{\odot}$ to 0.02 $\rm Z_{\odot}$.  
The scatter in the \zh-\zs\ relation also increases with  decreasing  metallicity. The $1\sigma$ ($68\%$) spread around \zsm\ varies from 0.4 dex to 0.7 dex between \zsm\ values of $\rm Z_{\odot}$ and 0.3 $\rm Z_{\odot}$ (or equivalently  between \zh\ values of 0.5 $\rm Z_{\odot}$ and 0.02 $\rm Z_{\odot}$).  

\begin{figure*}
\centering
\begin{tabular}{c}
\includegraphics[width=1.0 \textwidth]{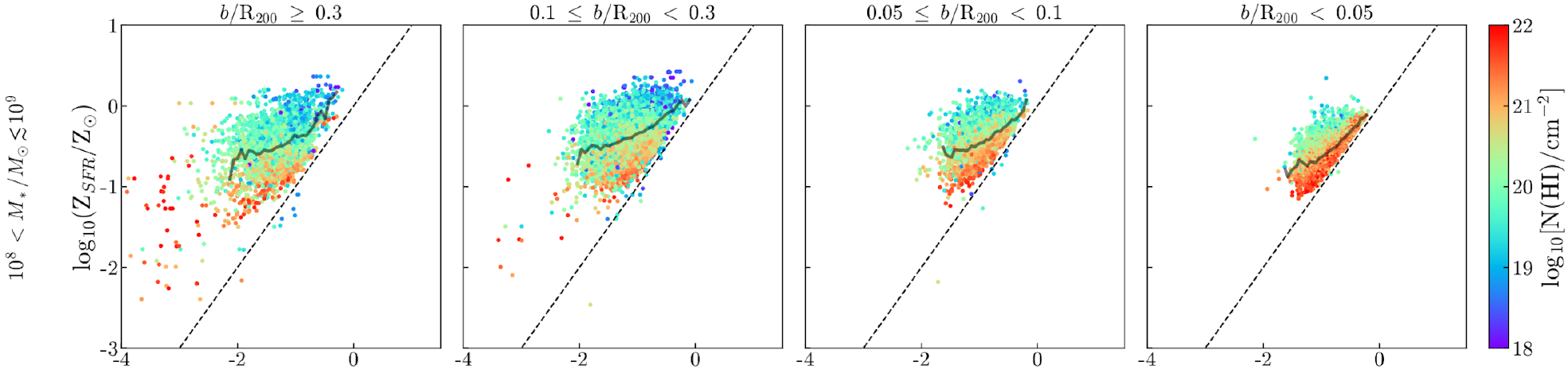} \cr
\includegraphics[width=1.0 \textwidth]{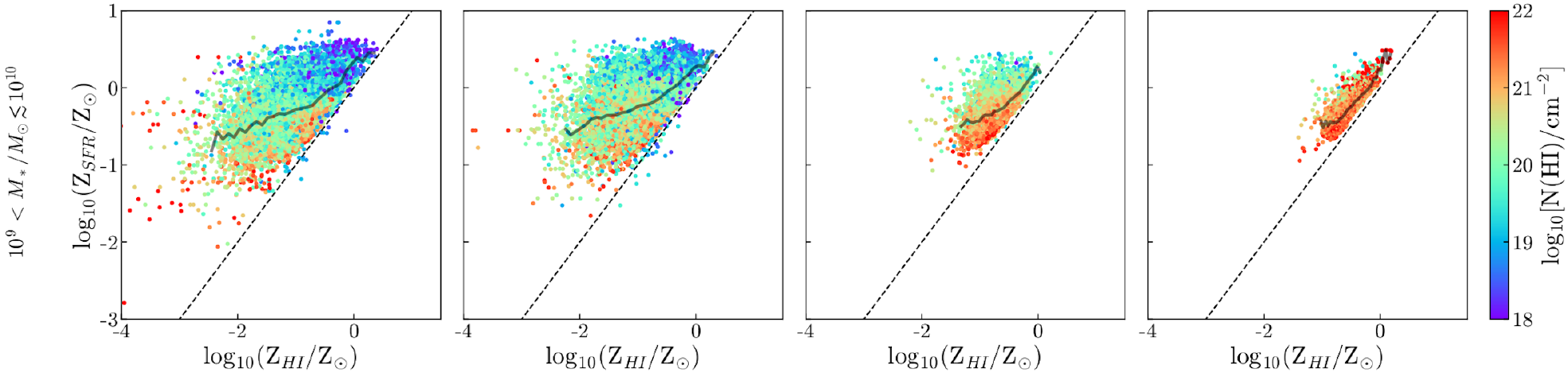} \cr
\includegraphics[width=1.0 \textwidth]{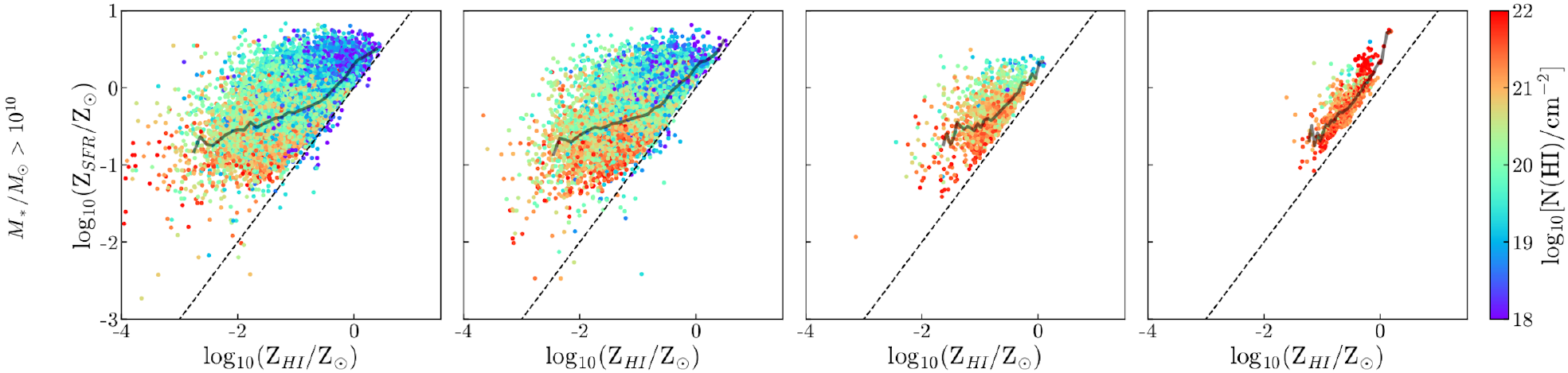}
\end{tabular}
\caption{The \zh--\zs\ plane for the sightlines at $z=2$ in three  bins of stellar mass (rows) and four bins of impact parameter (columns). The colour of the points indicates the N(\hi) value of each sightline, with the colour bar presented in the right side of each row. The grey lines show the $Z_{SFR,med}$ values in each plot. The x=y line in the plane is presented by a dashed line. 
Note that R$_{200}$ varies   between $\sim 40-90$ kpc for galaxies with stellar masses between $\rm 10^8$ and $\rm 10^{10}\,M_{\odot}$.  
\label{fig:9panels}}
\end{figure*}

The large differences between the metallicities of the two gas phases  in the  low metallicity regime indicates that the low metallicities of the atomic   gas could have a large contribution from dilution of heavy elements through mixing with the metal-poor, infalling pristine gas that is not yet enriched by heavy elements. 
{We note that the metals in the EAGLE simulation remain fixed to their Lagrangian resolution element after being injected by a stellar population. Therefore, the mixing we refer to here is not an exchange of metals between resolution elements. Here, `mixing' refers to the gas (resolution elements) moving in(to) and out of the galaxy, intermingling and transporting metals and pristine gas, or in  other words, it refers to the effective distribution of metals in the galaxy through multiple feedback events. 
The dilution by inflowing pristine gas is particularly expected in the outskirts of galaxies,}  
where the infall of gas from intergalactic medium reduces the average metal enrichment of the diffuse, \hi\ without directly affecting the metallicity of the ionised gas \citep[][]{Collacchioni20-2020MNRAS.495.2827C, Wright21-2021MNRAS.504.5702W}. 
On the other hand, the decreased difference between the two metallicities in the high-metallicity regime could also imply that the mixing of neutral and ionised gas  in metal-rich regions (such as the  central  regions of galaxies) is more efficient, leading  to a more uniform  metal enrichment in  different gas phases. The  enrichment of metals due to gas mixing in different phases,  and  the dilution of heavy elements in neutral gas   by mixing with  the  pristine gas, are both likely to contribute in the difference between the metallicities of atomic and ionised gas. Although, the contributions of these two effects seem to depend on, directly or indirectly,  the  metallicity regime. 

It is important to note that the sightlines with low metallicities in Figure \ref{fig:all-z2} not only  sample the  (metal-poor) outskirts of galaxies (independent of the stellar mass regime), but also sample the entire extent of the low-mass galaxies, irrespective of the positions of the  sightlines in galaxies  with respect to the galaxy centres   
\citep[given the mass-metallicity relation of star-forming galaxies;][]{Lequeux1979, Tremonti04-2004ApJ...613..898T, Zahid14-2014ApJ...791..130Z, Kudritzki15-2015MNRAS.450..342K}. 
The large  $\delta$ values in the low metallicity regime therefore raises the following questions: ``is the mixing of gas in 
low-mass galaxies less efficient than in massive galaxies?'',  ``does  the position of  a sightline   with respect to the centre of the galaxy  have a significant effect on the mixing of gas?'', and ``does the stellar mass of a galaxy  affect the contribution of the pristine gas in diluting the metal enrichment of \hi\ in the system and result in lower values of $\delta$?'',

To answer these questions, one  needs to separately investigate the  effects of the  stellar mass and the  impact parameter on the  \zh--\zs\ relation. 
Considering  that   lower-mass galaxies are  smaller in size \citep[see e.g.,][for the stellar-mass-size relation of galaxies]{Shen03-2003MNRAS.343..978S, Graham08-2008MNRAS.388.1708G, Maltby10-2010MNRAS.402..282M, vanderwel14-2014ApJ...788...28V}, they are expected to  host sightlines with generally  smaller impact parameters. We therefore use the  impact parameter of the sightlines, normalised by the R$_{200}$ values of galaxies (varying   between $\sim 40-90$ kpc for galaxies with stellar masses between $\rm 10^8$ and $\rm 10^{10}\,M_{\odot}$) in our analysis. 
The large values of $b/$R$_{200}$ therefore corresponds to the outskirts of galaxies, independent of the stellar mass or equivalently the size of a galaxy.

Figure \ref{fig:9panels} compares the  effects of the stellar mass and the impact parameter on the \zh--\zs\ relation at $z=2$.   
Each row in the figure represents a stellar mass bin, and each column represents a specific range of the normalised impact parameter. The colouring of the points are  according to the \hi\  column densities  along the sightlines. 
It is clear that at a given metallicity the sightlines with smaller normalised  impact parameters  are more concentrated around the x=y line. In other words, the difference between the \zh\ and \zs\ values reduces in the central regions of galaxies, irrespective of their stellar mass.  

The stellar mass of galaxies does not seem to have a direct impact on the relation between  \zh\ and \zs. Of course galaxies with higher stellar mass would have larger sizes and naturally would accommodate more sightlines at larger impact parameters. Therefore the increase in $\delta$ and also the scatter of the relation at larger impact parameters (see the column for $0.1 \leq b/$R$_{200} < 0.3$) in galaxies with higher stellar mass is expected. 
At small impact parameters (see the column for $b/$R$_{200} < 0.05$), both $\delta$ and the scatter in the \zh--\zs\ relation appears to decrease as the stellar mass of the host galaxies increases. But noting the smaller spread of  \hi\ column densities in massive galaxies in this regime of impact parameters, the smaller deviation between the two metallicities  is likely not directly due to larger stellar mass, but is plausibly linked to the column density of \hi.   
In fact, the smaller deviations between the two metallicities and the smaller scatter of the relation in low impact parameters in all stellar mass bins could be a result of higher column densities -- \hi\ column densities are generally higher in the central regions of galaxies.

\begin{figure*}
\centering
\begin{tabular}{cc}
\includegraphics[width=0.45 \textwidth]{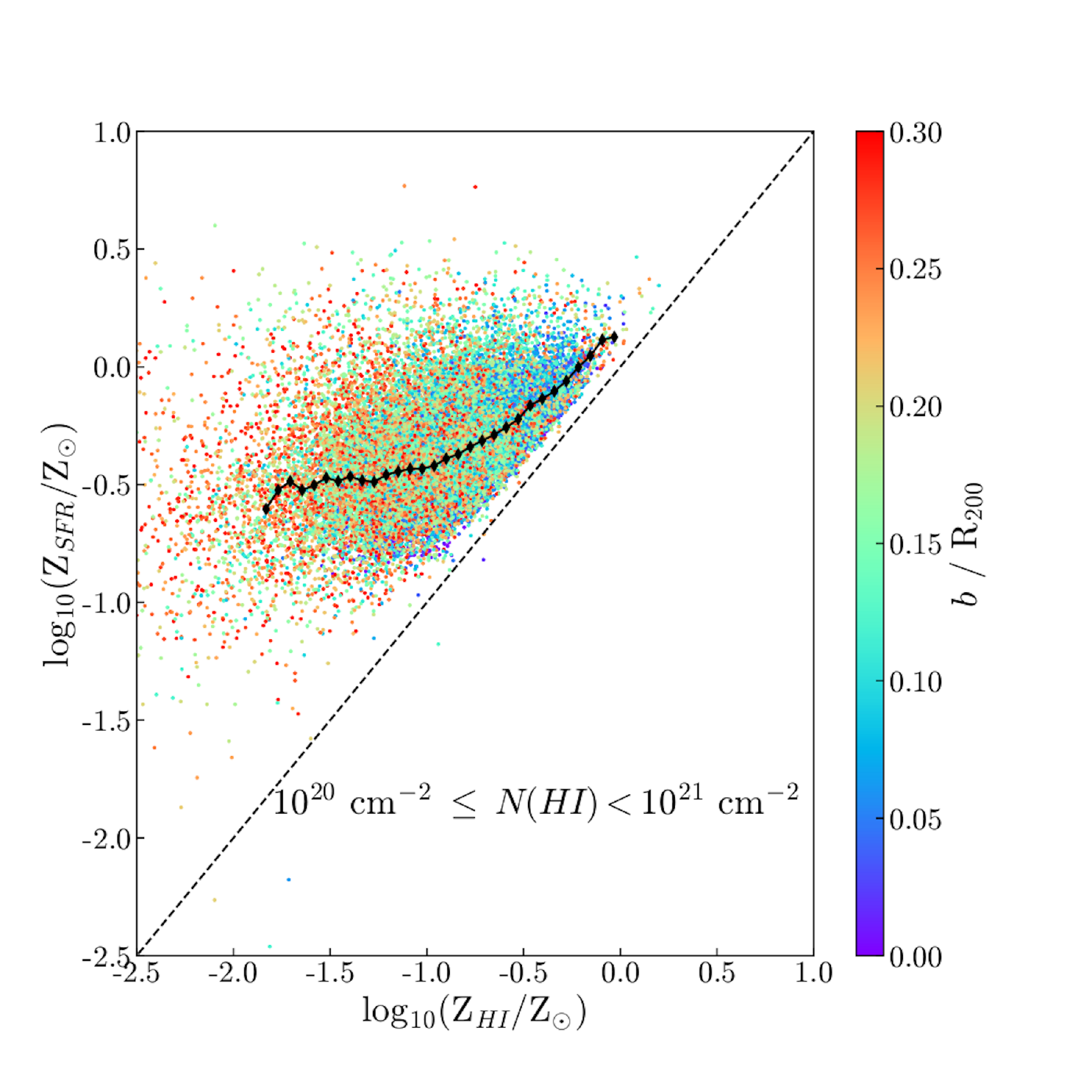} &
\includegraphics[width=0.45 \textwidth]{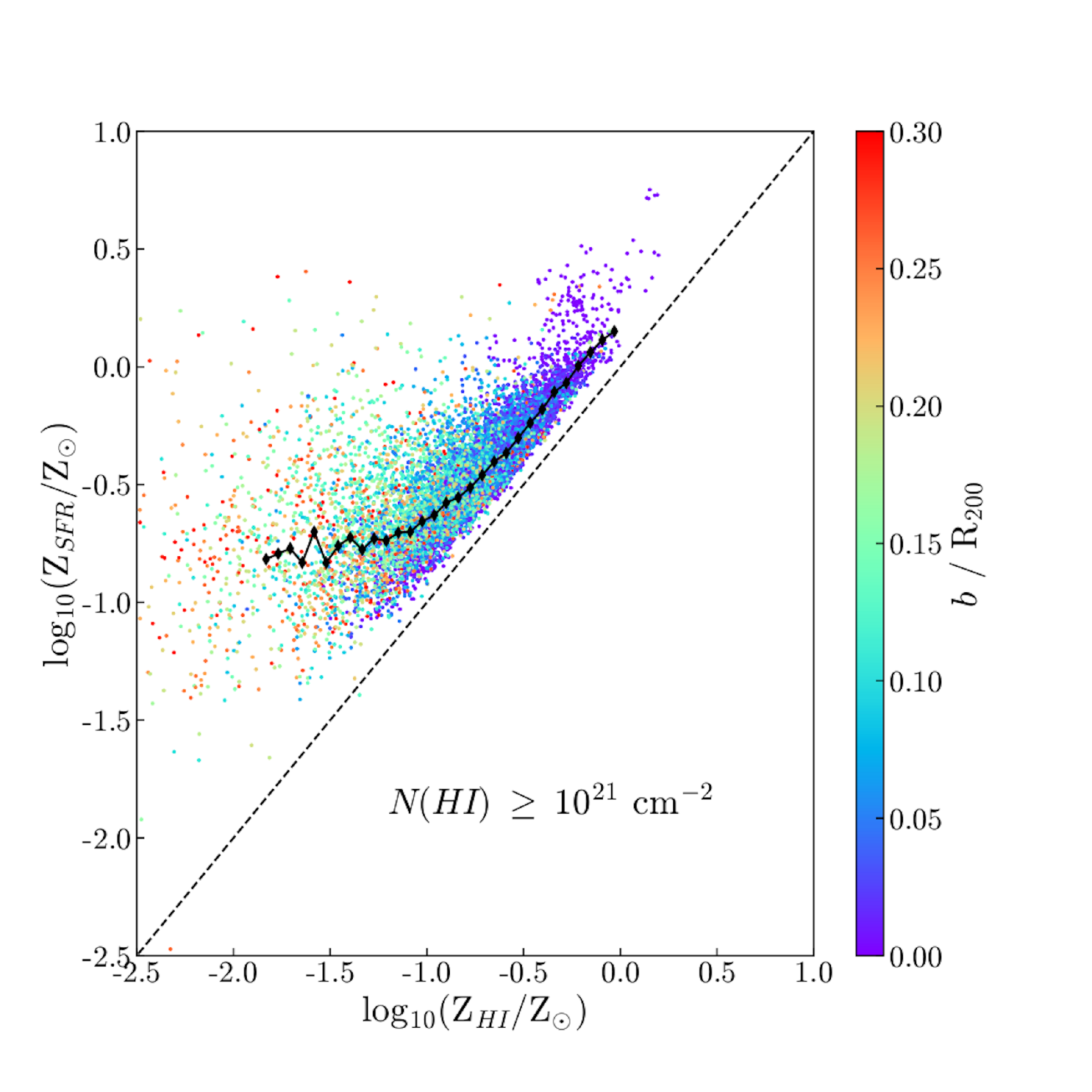} \cr
\includegraphics[width=0.45 \textwidth]{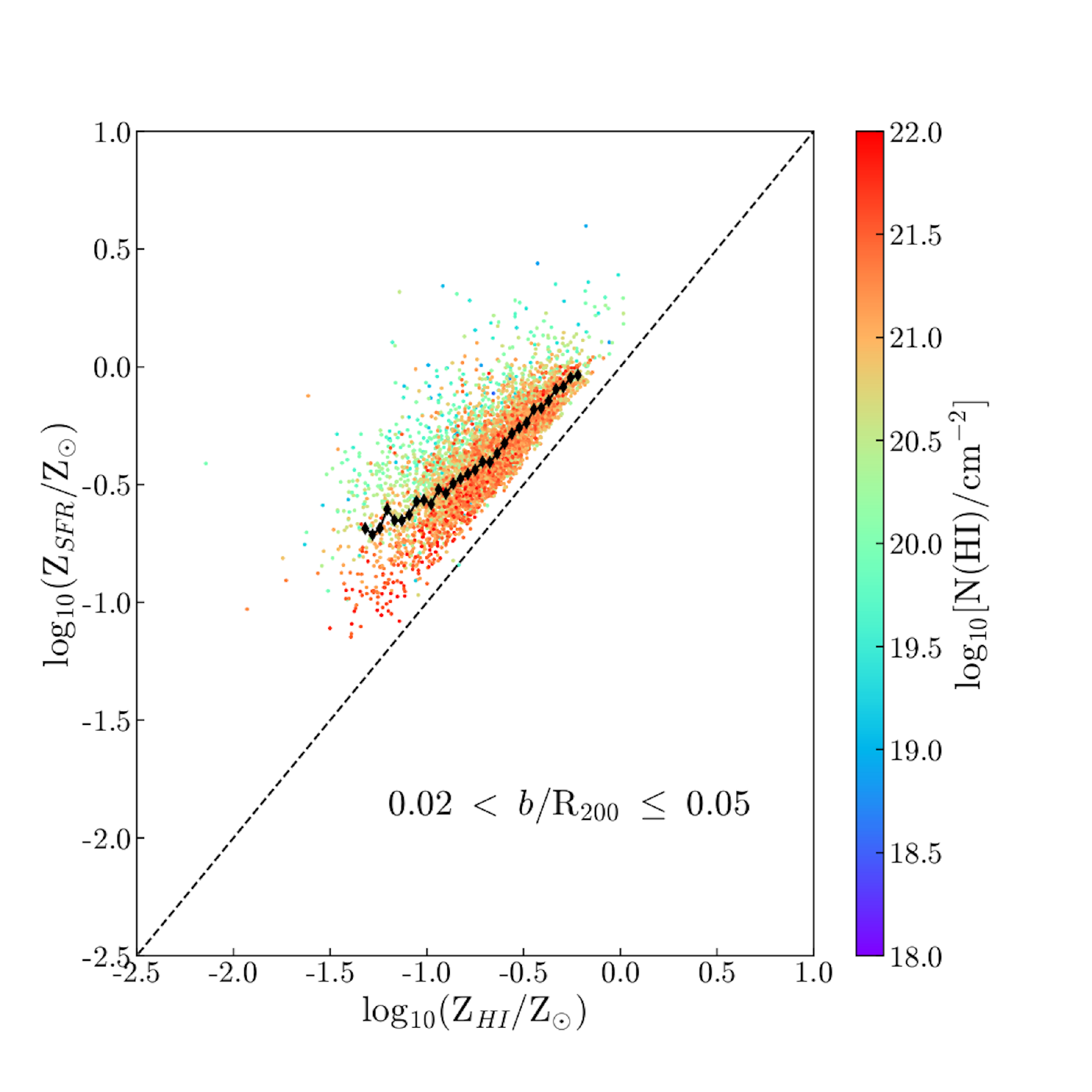} &
\includegraphics[width=0.45 \textwidth]{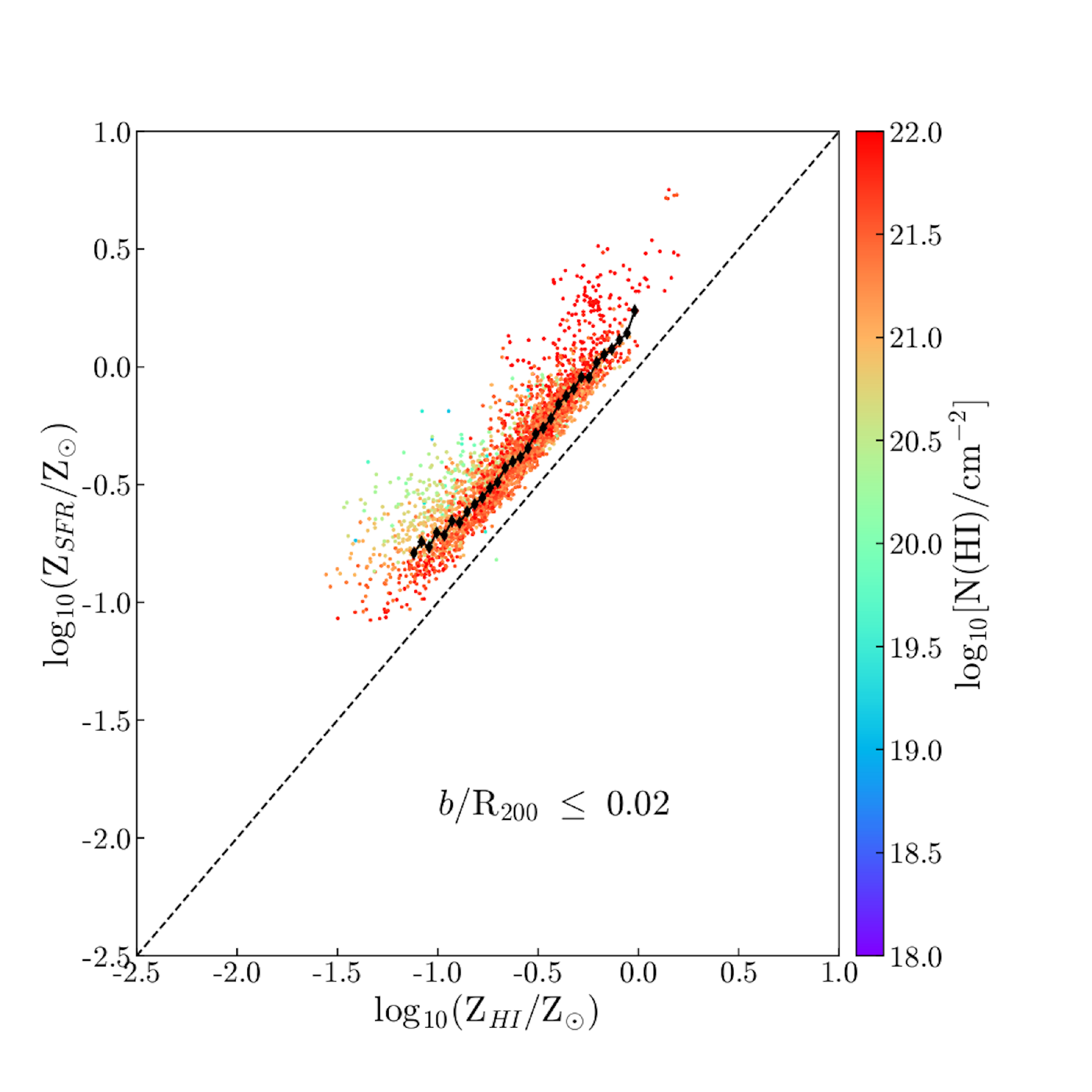} 
\end{tabular}
\caption{ 
The \zh--\zs\ plane for  sightlines at $z=2$ selected based on N(\hi) and impact parameter. The \zsm\ values are shown with black diamonds and the x=y line is presented with the dashed lines.  
{\bf Top panels:} Sightlines with \hi\ column densities of $\rm 10^{20}\ cm^{-2} \leq $ N(\hi) $\rm < 10^{21}\ cm^{-2}$ and N(\hi) $\rm\geq 10^{21}\ cm^{-2}$ in the left and right panels respectively. All the presented sightlines are within the 0.3 R$_{200}$ from galaxy centres. The colour bar  represents the value for  $b/$R$_{200}$.  
{\bf Bottom panels:} Sightlines with $0.02 < b/$R$_{200} \leq 0.05$  in left and those within the 0.02 R$_{200}$ central regions of galaxies  in the right panel respectively. The colour bar represents the \hi\ column density of the sightlines.    
\label{fig:4panels}} 
\end{figure*}

To investigate   the effect of the N(\hi) on $\delta$ and the scatter of the \zh--\zs\ relation, we  consider  only the sightlines with high \hi\ column densities -- N(\hi) $\rm \geq 10^{20} cm^{-2}$, the column density typically found within the optical disks of galaxies. We also restrict the sightlines to be within a  distance of $b/$R$_{200} \leq 0.3$ from the galaxy centres. The \zh-\zs\ plots for these sightlines at $z=2$ are presented in the top panels of Figure \ref{fig:4panels}. $\delta$ and the scatter in the  \zh-\zs\ relation remain quite large even at such high column densities. In fact, there is a significant fraction of sightlines with high \hi\ column densities for which the difference between the two metallicities exceeds 0.5 dex.

On the other hand, when we consider the sightlines within the 0.02 R$_{200}$ central regions of galaxies (irrespective of their \hi\  column densities), $\delta$ and the overall scatter on the \zh-\zs\ relation appear to be much smaller. The \zh-\zs\ plot for these sightlines are shown in the bottom-right panel of Figure \ref{fig:4panels}. It is clear from  this Figure that  these sightlines have large N(\hi) values, as expected to be the case in the central regions of galaxies.

Note that the region within 0.02 R$_{200}$ corresponds to the $\sim 1$ kpc, $\sim 1.5$ kpc, and $\sim 2$ kpc central regions of galaxies with stellar masses of $\rm 10^8\,M_{\odot}$, $\rm 10^9\,M_{\odot}$, and $\rm 10^{10}\,M_{\odot}$, respectively.
The gravitational softening length of 0.35 kpc in the simulations   thus allows us to explore  the separation of metallicity between gas phases at  spatial scales of 0.02 R$_{200}$.


\begin{figure*}
\centering
\begin{tabular}{cc}
\includegraphics[width=0.45 \textwidth]{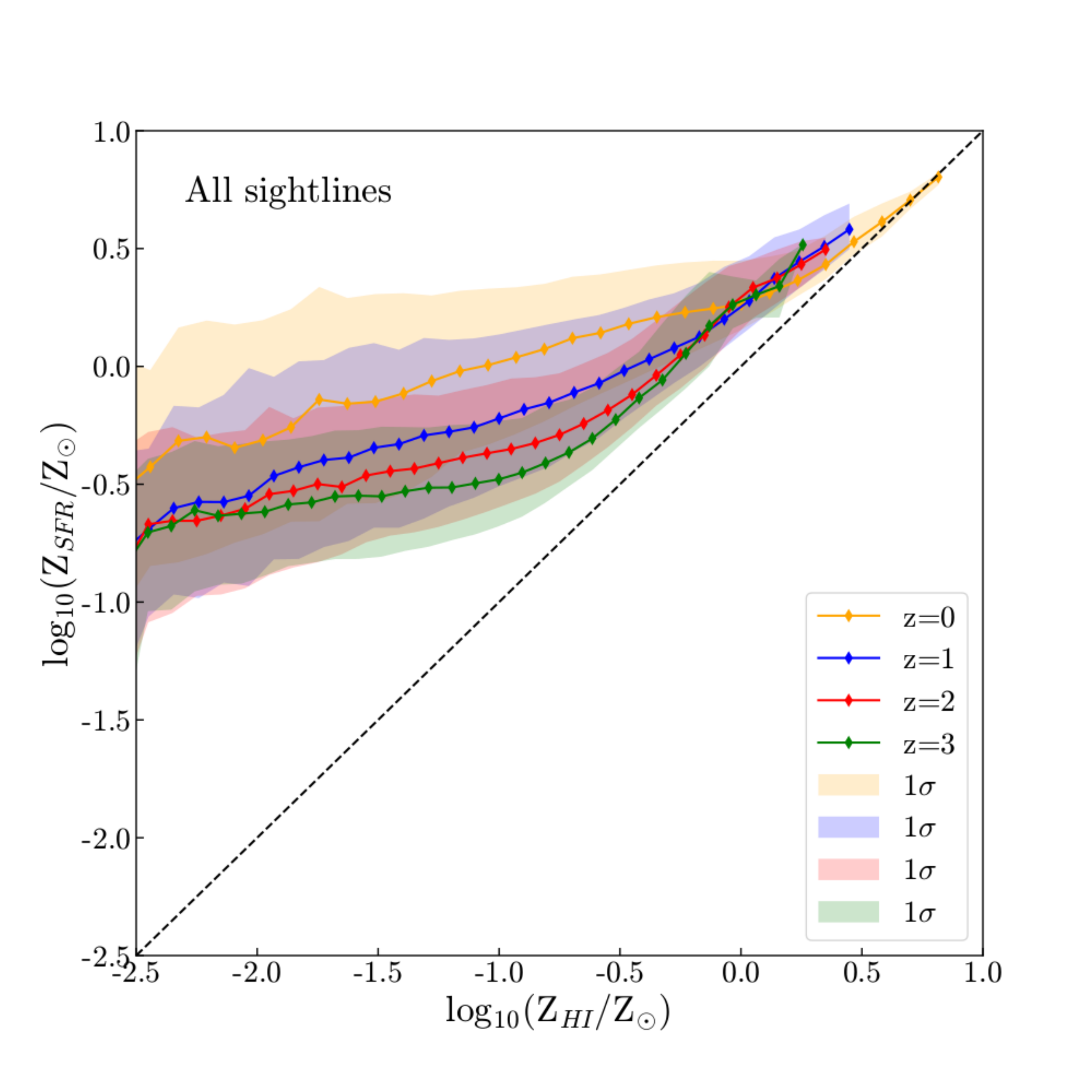} &
\includegraphics[width=0.45 \textwidth]{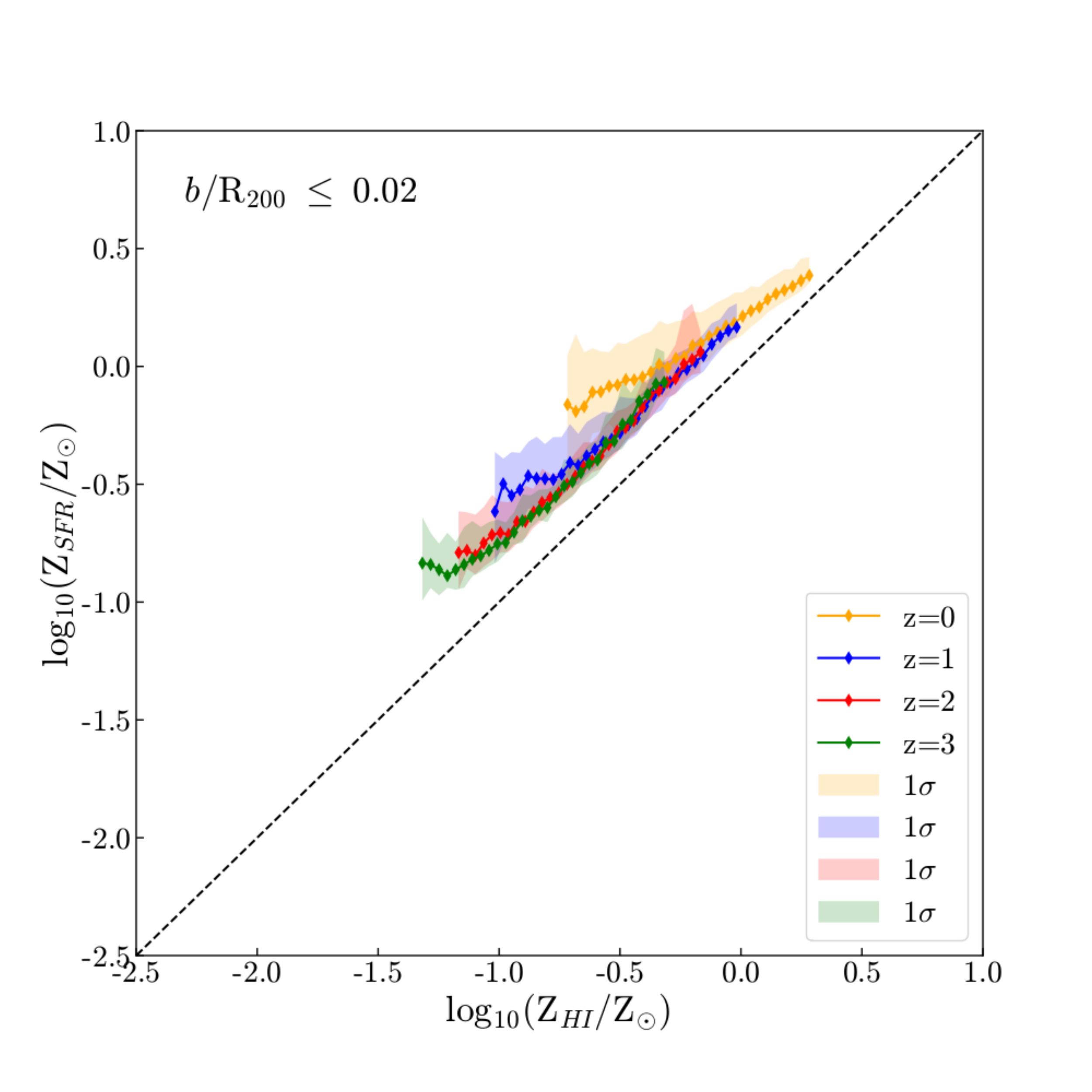} 
\end{tabular}
\caption{ 
{\bf Left:} The \zh-\zs\ relation for all the sightlines of EAGLE galaxies at {redshifts $z=0,1,2$ and 3}. The \zsm\ values are marked with diamonds and the $1\sigma$ scatter around the medians are shown with shaded areas. 
{\bf Right:} The \zsm\ values and the $1\sigma$ scatter around them for sightlines within the 0.02 R$_{200}$ central regions of galaxies {at} {four} redshifts.  
\label{fig:zevolution}}
\end{figure*}

In order to quantitatively compare  the role  of N(\hi) and the impact parameter on the difference between the two metallicities, we measure $\delta$ and the $1\sigma$ scatter of the \zh--\zs\ relation for the sightlines with  N(\hi) $\rm \gtrsim 10^{21}\ cm^{-2}$ (shown in the top-right panel of Figure \ref{fig:4panels})  and also the sightlines located within the 0.02 R$_{200}$ central regions of galaxies (presented in the bottom-right panel of Figure \ref{fig:4panels}). 
For the  \zsm\ values of $\gtrsim$ 0.3 $\rm Z_{\odot}$,  $\delta$  remains around 0.2 dex for both the groups, and the $1\sigma$ scatter around $\delta$ remains around 0.1-0.2 dex for both the groups as well. The \zh--\zs\ relation in this metallicity regime therefore is very similar for sightlines  with N(\hi) $\rm \gtrsim 10^{21}\ cm^{-2}$ and sightlines in the 0.02 R$_{200}$ central regions of galaxies.  However, both $\delta$ and the  $1\sigma$ scatter increase  rapidly  for the  former group at \zsm\ $\rm \lesssim 0.3 Z_{\odot}$.

We can therefore safely state that   the location of a sightline being within the 0.02 R$_{200}$ central regions of galaxies guarantees a small, systematic  difference between  \zh\ and \zs\  with $\delta=$ 0.2 dex. 
This finding implies  that the mixing between star-forming gas and the \hi\ gas is efficient in the central regions of galaxies {(the mixing time scale is shorter than the typical gas depletion time). Note that this is not a direct consequence of {a} subgrid model for mixing since the simulations do not model {subgrid}  diffusion.} The efficiency of mixing  seems to reduce  by moving outwards from the galaxy centres, as indicated by the large values of $\delta$ in large impact parameters. We find this to be  independent of the stellar mass of the galaxies and hence the metallicity regime, as discussed before (see Figure \ref{fig:4panels}). 

On the other hand, a high \hi\ column density of a sightline alone does not imply a small difference between the two metallicity measurements, and depending on the metallicity regime, $\delta$ could be as high as $\sim$ 1.5 dex  even for a sightline with   N(\hi) $\rm \gtrsim 10^{21}\ cm^{-2}$.  
Sightlines located at large distances from galaxy centres, even those  with high \hi\ column densities, are less enriched with heavy elements  compared to the ionised gas in projected  regions which are more metal enriched  due to  local star formation. 
We note that some of these sightlines could be associated with  low-mass satellite galaxies, and some could be linked to the  intergalactic medium, appearing at certain impact parameters due to projection effects.   
But given the significant number of high N(\hi) sightlines, with large $\delta$ values,  a large number of them should be associated with the outskirts of galaxies. 
Given the high N(\hi) of these sightlines, they are less likely to be  affected by mixing with pristine gas. Therefore, their large $\delta$ values should be mainly due to the inefficiency of the mixing of  star-forming gas with  \hi. This implies that the inefficiency of the mixing of the two gas phases  should be  the dominant effect compared to the metal dilution in \hi\ through  mixing with   pristine gas in the outskirts.
\citet[][]{Mitchell20-2020MNRAS.497.4495M} and \citet[][]{Wright21-2021MNRAS.504.5702W}  used the EAGLE simulations and showed that accreting gas in massive galaxies have more contributions from the recycled (hence more metal-rich) gas   compared to low mass galaxies. 
Given that the stellar mass has an  insignificant role on the \zh-\zs\ relation, the findings of \citet[][]{Mitchell20-2020MNRAS.497.4495M} and \citet[][]{Wright21-2021MNRAS.504.5702W}  support the dominant effect of the mixing of the two gas phases in the outskirts of galaxies.     

We also investigate the redshift evolution of the \zh--\zs\ relation {in the redshift range $z=0-3$}. The left panel of Figure \ref{fig:zevolution} shows the \zsm\ values and the $1\sigma$ scatter of the relation for all the sightlines at redshifts {$z=0,1,2$ and 3}. The \zh--\zs\ relation appears to have a slight evolution with redshift: the \zsm\ of sightlines at a fixed \zh\ slightly decreases with  increasing  redshift. This is consistent with the  overall decrease in the average ionised gas metallicity in galaxies  measured from observations.    However, the relation between the two metallicities does not  vary much with redshift for sightlines in the central regions of galaxies,  as demonstrated in the right panel of Figure \ref{fig:zevolution}. The small, systematic difference between  \zh\ and \zs\ in the central regions of galaxies remains around 0.2 dex  between  $z=0$ to $z=3$. This suggests  the efficiency of  gas mixing in the central regions of galaxies to be very similar  at $z=0-3$.

\subsection{Comparison with observations}

\begin{figure}
\centering
\includegraphics[width=0.5 \textwidth]{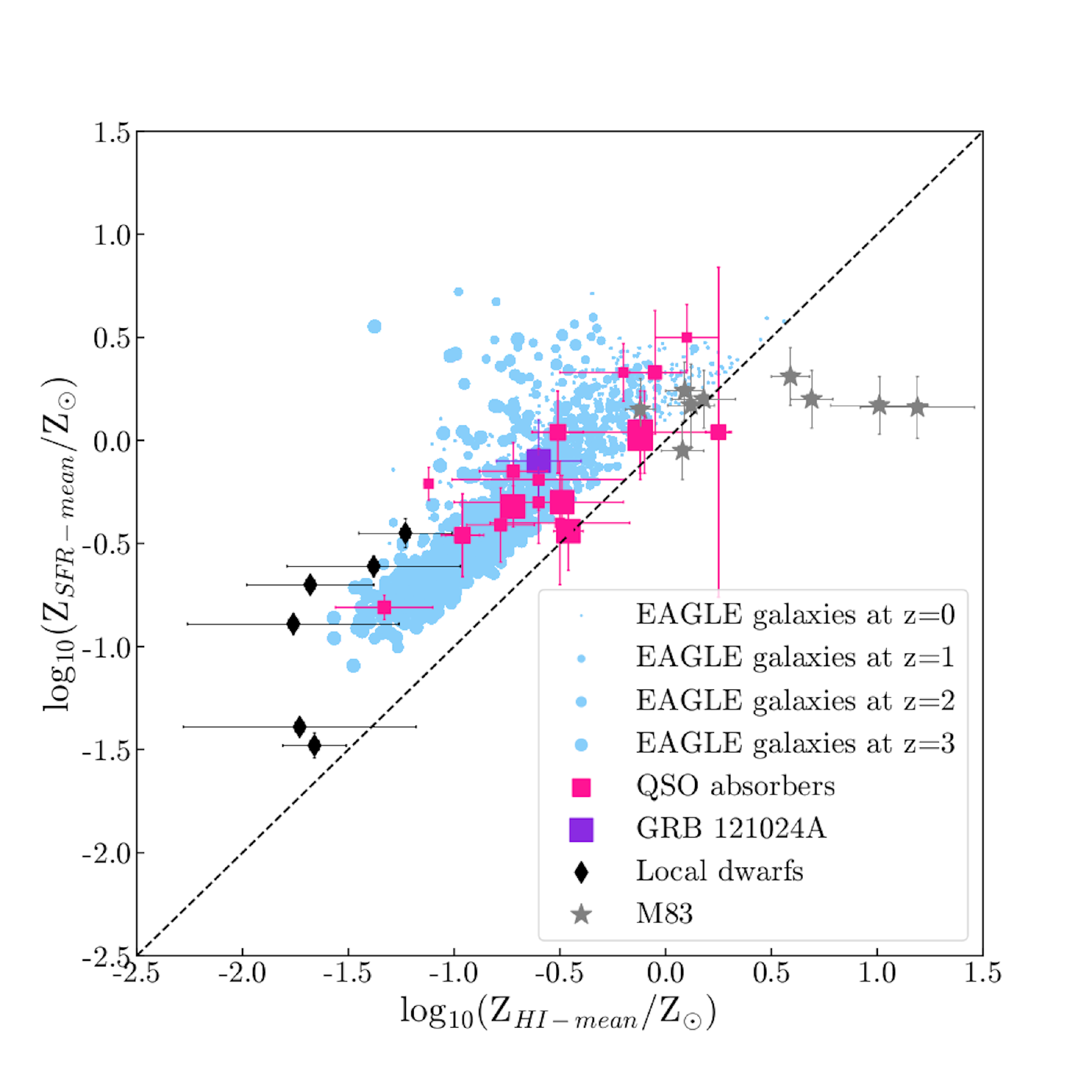}
\caption{ 
The ionised and atomic  gas metallicity measurements from observations, in comparison with the obtained metallicities of EAGLE galaxies. 
The quasar absorbers are shown with pink squares, and  the host galaxy of GRB 121024 is presented with a purple square. 
Nearby dwarf galaxies are marked by black diamonds. The individual regions of M83 are shown by gray stars. 
The x values for these systems are measured in  single sightline  from the ratio between the column densities of heavy elements and \hi, obtained from ISM absorption lines.  The y values for these systems are obtained from the flux measurement of the bright nebular lines coming from the entire star-forming disk of the galaxy. The size of the squares are proportional to the redshift of the system. 
The blue points show the EAGLE galaxies at $z=0-3$ in the $\rm Z_{HI,mean}$--$\rm Z_{SFR,mean}$ plane. 
\label{fig:obs}}
\end{figure}

To compare our findings  from simulations with the available observations, we use the  metallicity measurements  of both ionised and atomic phases of gas in the same galaxies reported in the literature. 
{These measurements and the relevant references are provided in Table \ref{tab:obs}. The gathered observational data are for  22 intervening galaxies along  quasar   lines of sight, a GRB host galaxy, and 7 nearby galaxies.} 

{For the nearby galaxies, the  metallicity  of \hi\ gas  is obtained from the ratio between the column densities of heavy elements (such as argon, nitrogen, and oxygen)  and \hi, derived  from the imprinted absorption lines  on the continuum emission form the bright UV targets (such as UV-bright massive
stars) within these galaxies. The reported atomic metallicities of these galaxies   in Table \ref{tab:obs} are all based on abundances of oxygen.  
For all these galaxies but M83 the values of  $\rm Z_{atomic}$ and $\rm Z_{ionised}$ are the average values over the entire galaxies. 
For galaxy M83, the reported metallicities  of  both gas phases are derived in  individual  regions in the galaxy. The impact parameters of all these regions are provided in Table. \ref{tab:obs}.}

The metallicities    of \hi\ in the QSO absorbers and the GRB host are derived based on the ratio between the column densities of heavy elements (such as Zn and S) and \hi, obtained from ISM absorption lines imprinted on  quasar/GRB spectrum. These metallicities are therefore measured for a  narrow beam along the sightlines of the bright background quasars or GRBs. The ionised gas metallicities on the other hand are derived from the flux measurement of the bright nebular lines, originating from the star-forming regions in galaxies. These metallicities therefore provide an average value  over the entire galaxies. 

For a fair comparison between the  measurements from observations and our findings from simulations,  we require a representative \zs\ value for each of the EAGLE galaxies which is measured over the entire star-forming disk of the galaxy, similar to the metallicity that is measured for ionised gas in galaxies based on observations. We obtain this value in two different ways: (i) {we measure the mean SFR-weighted metallicity of all star-forming particles within the stellar half-mass radius of the galaxy,} and (ii) we obtain the mean of the \zs\ values for all the sightlines in a galaxy (within a side-length of 2R$_{200}$, as discussed in Section \ref{sec:s}). We find the two mean metallicities to be in very good agreement, and therefore use the latter definition for $\rm Z_{SFR,mean}$  as the global  \zs\ in each of the EAGLE galaxies.  We also use the mean values of the \zh, {$\rm Z_{HI,mean}$, by measuring the mean \hi-weighted metallicity of all particles within a radial aperture of a galaxy. 

{All the systems in Table \ref{tab:obs} have a range of \hi\ column density between $\rm \sim  10^{19}\ cm^{-2}$ and $\rm 10^{22} cm^{-2}$. The QSO-sightlines and all the regions in M83   have  an impact parameter $\lesssim$ 60 kpc. For the  host galaxy of GRB 121024, the impact parameter of the GRB sightline in its host galaxy is not measured, but it is likely to be within the central few kpc of its host galaxy \citep[see][]{Friis15-2015MNRAS.451..167F}. The median impact parameter measured for GRBs is reported to be $\sim 1-2$  kpc \citep[see][]{Arabsalmani15-2015MNRAS.446..990A, Lyman17-2017MNRAS.467.1795L}. For the nearby dwarfs, given their sizes, the continuum emission {originates} from regions  within the $\lesssim$ 60 kpc range.  We therefore restrict the $\rm Z_{HI,mean}$ measurements to \hi\ column densities $\rm 10^{19}\ cm^{-2} - \rm 10^{22}\ cm^{-2}$ and impact parameters of 0-60 kpc for  a fair comparison with the observational measurements in hand.} 
Figure \ref{fig:obs} presents the metallicity measurements of  all the  systems from observations (a total number  of 37 data points), along with the obtained metallicities of EAGLE galaxies in the $\rm Z_{HI,mean}$--$\rm Z_{SFR,mean}$ plane,  {at $z=0-3$, similar to the  redshifts of the  observational points. 

We find a good agreement between the metallicity measurements from observations and the prediction from EAGLE simulations.  
The four observational data points that stand out in the relation correspond to the four regions in the 200 pc nuclear region of M83.
\citet[][]{Hernandez12-2021ApJ...908..226H} discusses that the efficient conversion of atomic to molecular gas and hence the depletion of the \hi\ gas in the nuclear  region of M83 is responsible for the {relatively high} atomic gas metallicity in that region. Our findings (see Fig. \ref{fig:9panels} and Fig. \ref{fig:4panels}) clearly demonstrate that simply having low \hi\ column densities in star forming regions (with normal mode of star formation) do not necessarily lead to \zs $<$ \zh. However, the presence of an  ongoing starburst in the nuclear region of M83   \citep[][]{Dopita10-2010ApJ...710..964D, Wofford11-2011ApJ...727..100W} might be the cause for the significant difference between the metallicities with \zs $<$ \zh.  This indicates that regions with a {nuclear} starburst mode of star formation might be following a different \zh-\zs\ relation. 
{Note that the effects of enhanced radiation fields in nuclear starbursts is not included when simulating the EAGLE galaxies.}

We also note that  a few of the local dwarf galaxies presented in Fig. \ref{fig:obs} have stellar masses $\sim \rm 10^6\,M_{\odot}$,  below the lower  stellar mass limit of the EAGLE galaxies ($\rm M_*=10^8\,M_{\odot}$). These dwarf galaxies trace the lower metallicity end of  the \zh-\zs\, relation predicted by the EAGLE galaxies.}

\begin{table*}
\centering
\caption{Systems with measured ionised and atomic gas metallicities, gathered from literature.}
\label{tab:gmcs}
\begin{tabular}{lcccccc}
\toprule
\toprule
Name   &  $z$ & $\rm log_{10}(N(HI)/cm^{-2})$ & $b$ (kpc) & $\rm Z_{atomic}/Z_{\odot}$ & $\rm Z_{ionised}/Z_{\odot}$ & References \\
\midrule
\midrule
\multicolumn{7}{c}{QSO-absorbers}\\
\midrule
Q0302-223    &    1.009	&	20.36	&	25	&	$-$0.51$\pm$0.12 & 	 0.04$\pm$0.20  &1,2	 \\
Q0452-1640   &    1.007	&	20.98	&	16	&	$-$0.96$\pm$0.08 &	$-$0.46$\pm$0.20	&2,3 \\
Q1009-0026   &    0.887 &	19.48	&	39	&	 0.25$\pm$0.06 &	 0.04$\pm$0.8	&1,2 \\
J1422-0001   &    0.91	&	20.40	&	12	&	$-$0.1$\pm$0.4   &	 0.04$\pm$0.2	&4	 \\
J1659+3735   &    0.1998&	18.89	&	58 	&	$-$1.12$\pm$0.02 &	$-$0.21$\pm$0.08	&5 \\
J2222-0946   &	2.354	&	20.65	&	6.3	&	$-$0.49$\pm$0.05	&	$-$0.30$\pm$0.13	&2,3\\
HE2243-60    &    2.324	&	20.60	&	26	&	$-$0.72$\pm$0.05 &	$-$0.32$\pm$0.10	&6 \\
Q0235+164ID2&	0.5243	&	21.70	&	13.2	&	$-$0.60$\pm$0.41	&	$-$0.19$\pm$0.15	&7\\
Q0439-433    &	0.101   &	19.85	&	7.2	&	$-$0.20$\pm$0.30	&	0.33$\pm$0.14	&7\\
Q0918+1636   &	2.5832	&	20.96	&	16.2	&	$-$0.12$\pm$0.05	&	0.01$\pm$0.20	&7\\
Q2222-0946   &	2.354	&	20.65	&	6.3	&	$-$0.46$\pm$0.07	&	$-$0.44$\pm$0.19	&7\\
J0238+1636   &	0.5253	&	21.70	&	7	&	$-$0.6$\pm$0.4	&	$-$0.3$\pm$0.2	&8\\
J0304-2211   &	1.0095	&	20.36	&	25	&	$-$0.51$\pm$0.12	&	0.04$\pm$0.2	&8\\
J0441-4313   &	0.1010	&	19.63	&	7	&	0.10$\pm$0.15	&	0.50$\pm$0.16	&8\\
J0452-1640   &	1.007	&	20.98	&	16	&	$-$0.96$\pm$0.10	&	$-$0.46$\pm$0.20	&8\\
J0918+1636   &	2.583	&	20.96	&	16	&	$-$0.12$\pm$0.05	&	0.04$\pm$0.2	&8\\
J0958+0549   &	0.6546	&	20.54	&	20	&	$-$1.33$\pm$0.23	&	$-$0.81$\pm$0.06	&8\\
J1009-0026   &	0.8864	&	19.48	&	39	&	 0.25$\pm$0.06	&	0.04$\pm$0.8	&8\\
J1138+0139   &	0.6126	&	21.25	&	12	&	$-$0.78$\pm$0.16	&	$-$0.41$\pm$0.18	&8\\
J1204+0953   &	0.6390	&	21.04	&	10	&	$-$0.72$\pm$0.16	&	$-$0.15$\pm$0.14	&8\\
J1436-0051   &	0.7390	&	20.08	&	50	&	$-$0.05$\pm$0.12	&	0.33$\pm$0.30	&8\\
J1544+5912   &	0.0102	&	20.35	&	1	&	$-$0.50$\pm$0.33	&	$-$0.4$\pm$0.30	&8\\
\midrule
\midrule
\multicolumn{7}{c}{GRB host galaxies}\\
\midrule
GRB 121024A	&	2.30	&	21.88	&		&	$-$0.6$\pm$0.2&	$-$0.1$\pm$0.2 &9\\
\midrule
\midrule
\multicolumn{7}{c}{M83}\\
\midrule
M83-3&&18.94&	0.2&	 1.19	$\pm$0.27	&0.16	$\pm$0.15&10\\
M83-4&&19.55&	0.2&     1.01	$\pm$0.23	&0.17	$\pm$0.14&10\\
M83-6&&20.72&	3.1&     0.12	$\pm$0.11	&0.17	$\pm$0.20&10\\
M83-8&&20.65&	3.8&     0.09	$\pm$0.09	&0.24	$\pm$0.14&10\\
M83-9&&20.56&	5.1&     0.08	$\pm$0.1	&$-$0.05	$\pm$0.14&10\\
M83-12&&20.98&	3.3&     $-$0.12	$\pm$0.07	&0.15	$\pm$0.15&10\\
M83-14&&20.43&	3.2&     0.18	$\pm$0.15	&0.20	$\pm$0.14&10\\
M83-POS1&&19.92&	0.2&     0.59	$\pm$0.09	&0.31	$\pm$0.14&10\\
M83-POS2&&18.91&	0.2&     0.69	$\pm$0.10	&0.20	$\pm$0.14&10\\
\midrule
\midrule
\multicolumn{7}{c}{Local dwarfs}\\
\midrule
POX36		&&20.28&&$-$1.38	$\pm$0.41	&$-$0.61	$\pm$0.05&11\\	 		
IZW18		&&21.34&&$-$1.66	$\pm$0.15	&$-$1.48	$\pm$0.06&12\\	 
SBS0335–052	&&21.86&&$-$1.73	$\pm$0.55	&$-$1.39	$\pm$0.01&13\\	 
IZw36		&&21.30&&$-$1.76	$\pm$0.50	&$-$0.89	$\pm$0.05&14\\	 
Mark59		&&20.85&&$-$1.68	$\pm$0.3	&$-$0.7	$\pm$0.003&15\\	 
NGC1705	        &&20.18&&$-$1.23	$\pm$0.22	&$-$0.45	$\pm$0.07&16\\	 
\hline
\bottomrule
\end{tabular}
\flushleft
{Columns: (1) name of the system, (2) redshift, (3) \hi\, column density, (4) impact parameter, (5) metallicity of atomic gas, (6) metallicity of ionised gas, (7) references. \\
References: (1) \citet[][]{Peroux11-2011MNRAS.410.2237P}; (2) \citet[][]{Augustin18-2018MNRAS.478.3120A}; (3) \citet{Peroux12-2012MNRAS.419.3060P}; (4) \citet{Bouche16-2016ApJ...820..121B}; (5) \citet[][]{Kacprzak14-2014ApJ...792L..12K}; (6) \citet{Bouche13-2013Sci...341...50B}; (7) \citet{Christensen14-2014MNRAS.445..225C} and references therein; (8) \citet{Rahmani16-2016MNRAS.463..980R} and references therein; (9) \citet{Friis15-2015MNRAS.451..167F}; (10) \citet[][]{Hernandez12-2021ApJ...908..226H}; (11) \citet[][]{Lebouteiller-2009A&A...494..915L}; (12) \citet[][]{Lebouteiller13-2013A&A...553A..16L}; (13) \citet{Thuan05-2005ApJ...621..269T}; (14) \citet{Lebouteiller04-2004A&A...415...55L}; (5) \citet{Thuan02-2002ApJ...565..941T}; (16) \citet{Heckman01-2001ApJ...554.1021H}.    
} 
\label{tab:obs} 
\end{table*}

\section{{discussion and Summary}}
\label{sec:sum}

The metal enrichment of   different gas phases in galaxies   can provide valuable information on fundamental phenomena in galaxy formation and evolution.  We present a detail study of the relation between the metallicities of ionised and atomic gas in star-forming galaxies at $z=0-3$ using the EAGLE simulations. For each of the simulated galaxies we construct several sight lines and for each sightline, we obtain the SFR- and N(\hi)- weighted metallicities of gas, \zs\ and \zh\ respectively, and use them  as proxies for the metallicities of ionised and atomic gas in the sightline. 

We find that  \zs\ $>$ \zh\ in almost all  sightlines, with $\delta$:=$\rm log_{10}(Z_{SFR,med}/Z_{\odot})$  exceeding a few dex in some sightlines, where $\rm Z_{SFR,med}$ is the median of \zs\ values in a \zh\ bin. 
In addition, $\delta$ increases with decreasing metallicity. 
{We investigate whether  stellar mass and hence the overall galaxy  metallicity   play a role  in gas mixing   and making the   metal enrichment in  different gas phases more uniform. Moreover we explore the effect of the impact parameter on $\delta$, considering that galaxy outskirts are less enriched with heavy elements and hence   the position of the sightlines with respect to the galaxy centre could have an   effect on   $\delta$ (e.g. through the  dilution of heavy elements by pristine gas in outskirts).} We find   that irrespective of galaxy  stellar mass,  $\delta$ decreases by decreasing the  impact parameter and reaches a minimum  value of $\sim$ 0.2 dex in the central regions of galaxies. 
This implies that mixing of star-forming gas and \hi\  is most efficient in the central region of galaxies. The large values of $\delta$ in large impact parameters indicate that the efficiency of mixing decreases by moving outwards from the galaxy centre. 
This finding is  independent of redshift, ie. $\delta$ saturates to $\sim$ 0.2 in the central 0.02 R$_{200}$ of galaxies at $z=0-3$.

Even though the  low impact parameter sightlines are generally associated with high   \hi\ column densities, having a high N(\hi) alone does not mean a small $\delta$. Atomic gas located at large distances from galaxy centres (large impact parameters), even those  with high \hi\ column densities, are less enriched with heavy elements  compared to the ionised gas in projected  regions which are more metal enriched  due to  local star formation.
In fact, we find  a significant number of sightlines with N(\hi) $\geq 10^{20}$ cm$^{-2}$ to be  associated with large  impact parameters, and    large values of $\delta$. Some of these sightlines could be associated with  low-mass satellite galaxies, and some could be linked to the  intergalactic medium, appearing at certain impact parameters due to projection effects. But  a large number of them are associated with the outskirts of galaxies. This suggests that in the galaxy outskirts, the large $\delta$ values  are mainly due to the inefficiency of mixing of star-forming gas with \hi\      rather than    dilution of heavy elements in \hi\   through mixing with the pristine gas.

\citet[][]{Mitchell20-2020MNRAS.497.4495M} and \citet[][]{Wright21-2021MNRAS.504.5702W}  used the EAGLE simulations and showed that accreting gas in massive galaxies have more contributions from the recycled (hence more metal-rich) gas   compared to low mass galaxies. 
Given that the stellar mass has an  insignificant role on the \zh-\zs\ relation, the findings of \citet[][]{Mitchell20-2020MNRAS.497.4495M} and \citet[][]{Wright21-2021MNRAS.504.5702W}  support the dominant effect of the mixing of the two gas phases in the outskirts of galaxies.     

We use the metallicity measurements of the ionised and atomic gas from observations, available in the literature, and compare them with the simulations. We find these observational measurements to  be in a good agreement with the \zh-\zs\ relation of galaxies in the EAGLE simulations. There is however an indication that  systems with startburst mode of star formation are likely to follow a different relation. Further analysis is required to investigate whether and how the presence of starburst can affect the metal enrichment of different gas phases. 

Increasing the number of observational data points is essential for drawing conclusions on the relation between the metallicity of the atomic and ionised gas phases in galaxies. 
Although there are a very large number of intervening systems detected in quasar sightlines with accurate atomic   gas metallicities, the presence of the bright background quasars make the ionised gas metallicity measurements quite challenging. 
Additionally,  quasar sightlines  typically trace the  gas in the outskirts of the intervening systems, located  at large distances ($\gtrsim 10$ kpc) from the star-forming regions where the ionised gas resides, making it even more challenging to measure the metallicity of the ionised gas in the same sightline. 
GRB sightlines on the other hand originate within   the star-forming regions of their host galaxies.  Therefore,  the metallicities  measured   from the bright  emission lines from the host galaxies  are more likely to be representative of the metallicity of ionised gas in the GRBs sightlines. This provides a better chance to have a fair comparison between the metallicity of the ionised gas in GRBs sightlines (even without  direct measurements) with the atomic gas metallicity  obtained from   imprinted absorption lines on GRBs spectra. Moreover, GRBs are transients and hence they do not interfere with the observations of the bright emission lines from their host galaxies after they fade away. 
Observations with James Webb Space Telescope (JWST) will allow increasing the number of GRB hosts with metallicity measurements from both the methods \citep[][]{Schady21-2021jwst.prop.2344S}. This, however, is limited by the number of detected GRB host galaxies.

Detailed studies of individual galaxies in sub-kpc scales, such as the study presented in \citet[][]{Hernandez12-2021ApJ...908..226H},   are proposed to be carried out by  combining the Integral Field Units (IFU) observations of galaxies with observations of bright background sources,  to derive the metallicities of both the gas phases in similar regions of nearby galaxies  \citep[see][]{Kulkarni2020-2020hst..prop16242K, James2020-2020hst..prop16240J}. Extending such studies to a large number of galaxies will 
allow us to test the  findings presented in this paper
, predicted by EAGLE, and will bring invaluable insights in our understanding of metal enrichment in galaxies.


\section*{Acknowledgments}
M.A. thanks   Vianney Lebouteiller, Martin A. Zwaan, and Lise Christensen  for helpful discussions.  
The presented study is funded by the Deutsche Forschungsgemeinschaft (DFG, German Research Foundation) under Germany´s Excellence Strategy – EXC-2094 – 390783311. 
We acknowledge the Virgo Consortium for making their simulation data available. The EAGLE simulations were performed using the DiRAC-2 facility at Durham, managed by the ICC, and the PRACE facility Curie based in France at TGCC, CEA, Bruy\`eres-le-Ch\^atel. 
This work was supported by resources provided by the Pawsey Supercomputing Centre with funding from the Australian Government and the Government of Western Australia.  
D.O. is a recipient of an Australian Research Council Future Fellowship (FT190100083) funded by the Australian Government. 
Nastasha Wijers was supported by a CIERA Postdoctoral Fellowship.




\appendix 

In the main body of the paper, we have used  the metallicities associated with each particle, as opposed to the SPH-smoothed version of the metallicities. Here we present  all the plots presented in Section \ref{sec:res} using the SPH-smoothed version of the particle metallicities. 

\begin{figure}
\centering
\includegraphics[width=0.5 \textwidth]{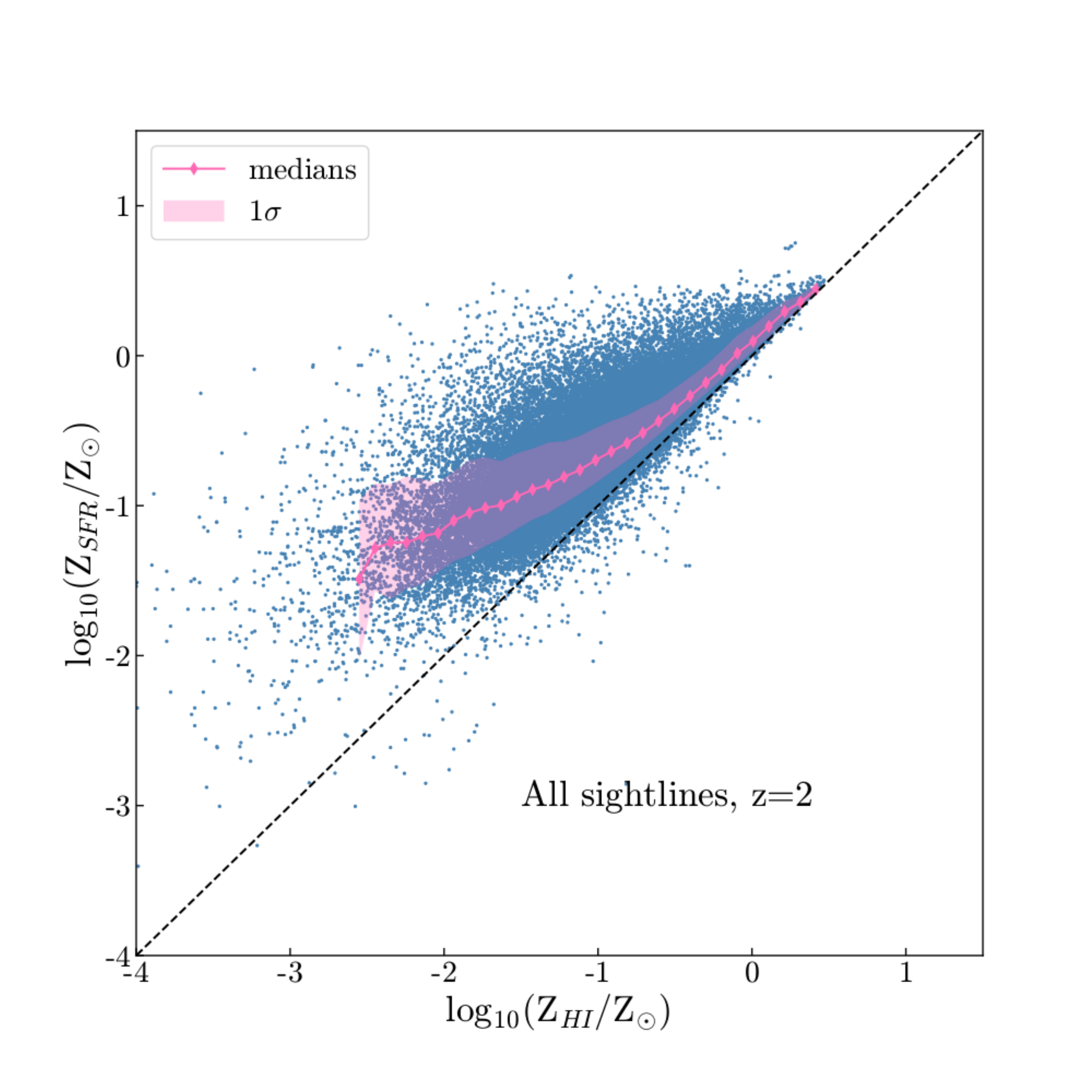}
\caption{ 
The \zh--\zs\ plane for all the sightlines in EAGLE galaxies at $z=2$ with smoothed metallicities. The medians of the \zs\ values in each \zh\ bin are shown with pink diamonds. The shaded area shows the  $1\sigma$ spread around the medians of the \zs\ values. The x=y line in the plane is presented by a dashed line.    
\label{fig:all-z2-smoothed}}
\end{figure}

\begin{figure*}
\centering
\begin{tabular}{c}
\includegraphics[width=1.0 \textwidth]{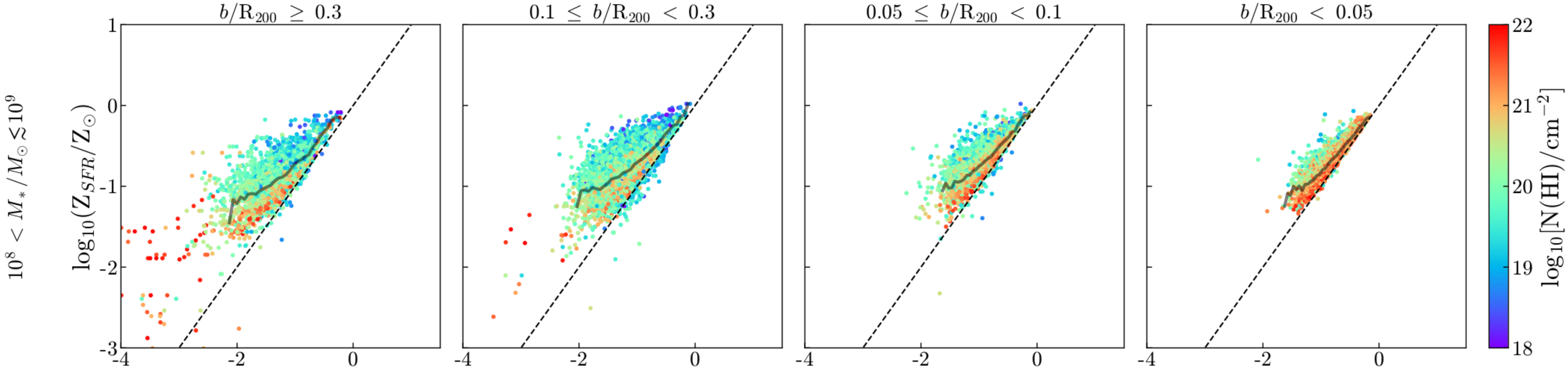} \cr
\includegraphics[width=1.0 \textwidth]{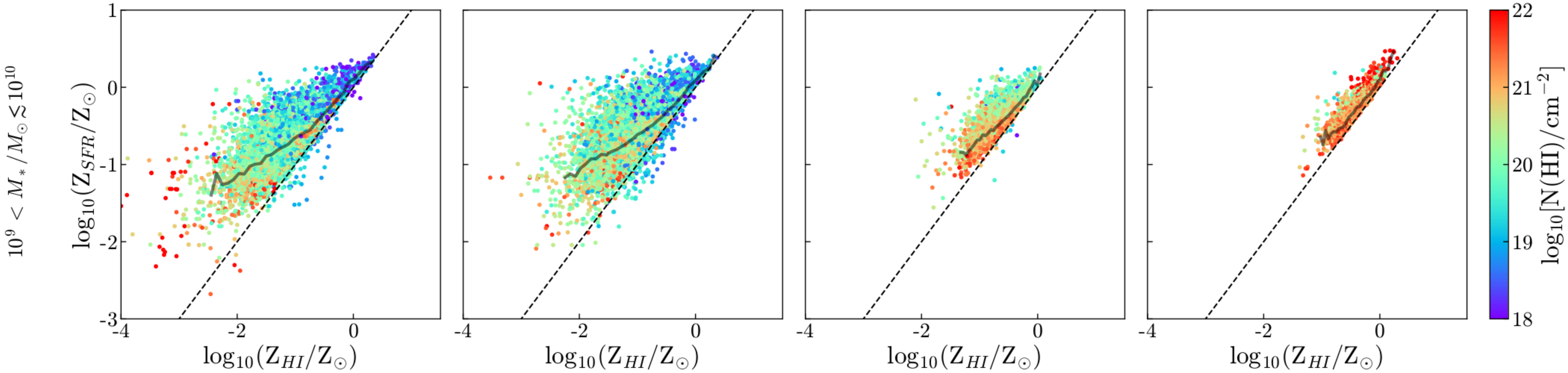} \cr
\includegraphics[width=1.0 \textwidth]{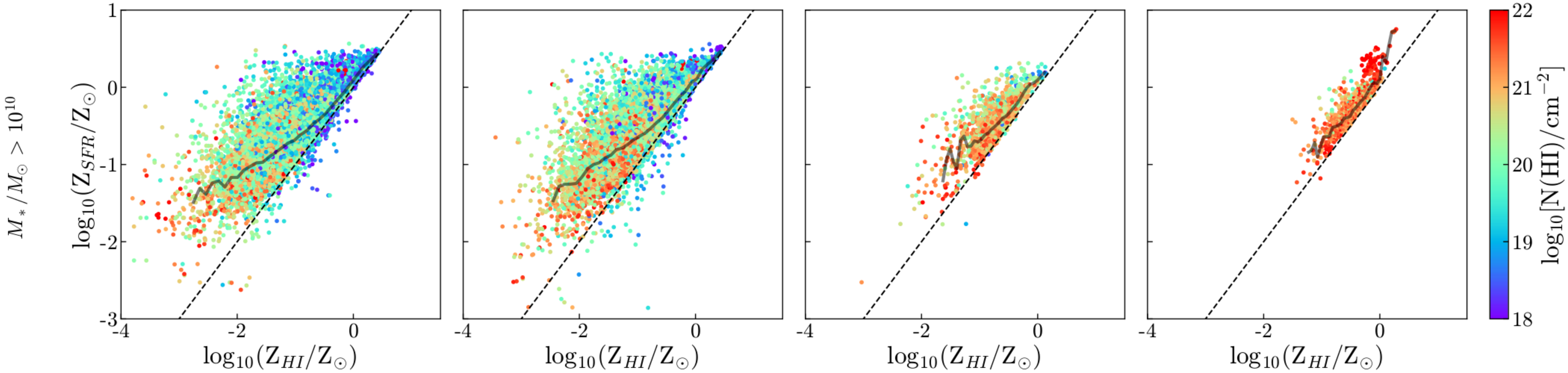}
\end{tabular}
\caption{The \zh--\zs\ plane with smoothed metallicities for  the sightlines at $z=2$ in three  bins of stellar mass (rows) and four bins of impact parameter (columns). The colour of the points indicates the N(\hi) value of each sightline, with the colour bar presented in the right side of each row. The grey lines show the $Z_{SFR,med}$ values in each plot. The x=y line in the plane is presented by a dashed line. 
Note that R$_{200}$ varies   between $\sim 40-90$ kpc for galaxies with stellar masses between $\rm 10^8$ and $\rm 10^{10}\,M_{\odot}$.  
\label{fig:9panels-smoothed}}
\end{figure*}

\begin{figure*}
\centering
\begin{tabular}{cc}
\includegraphics[width=0.45 \textwidth]{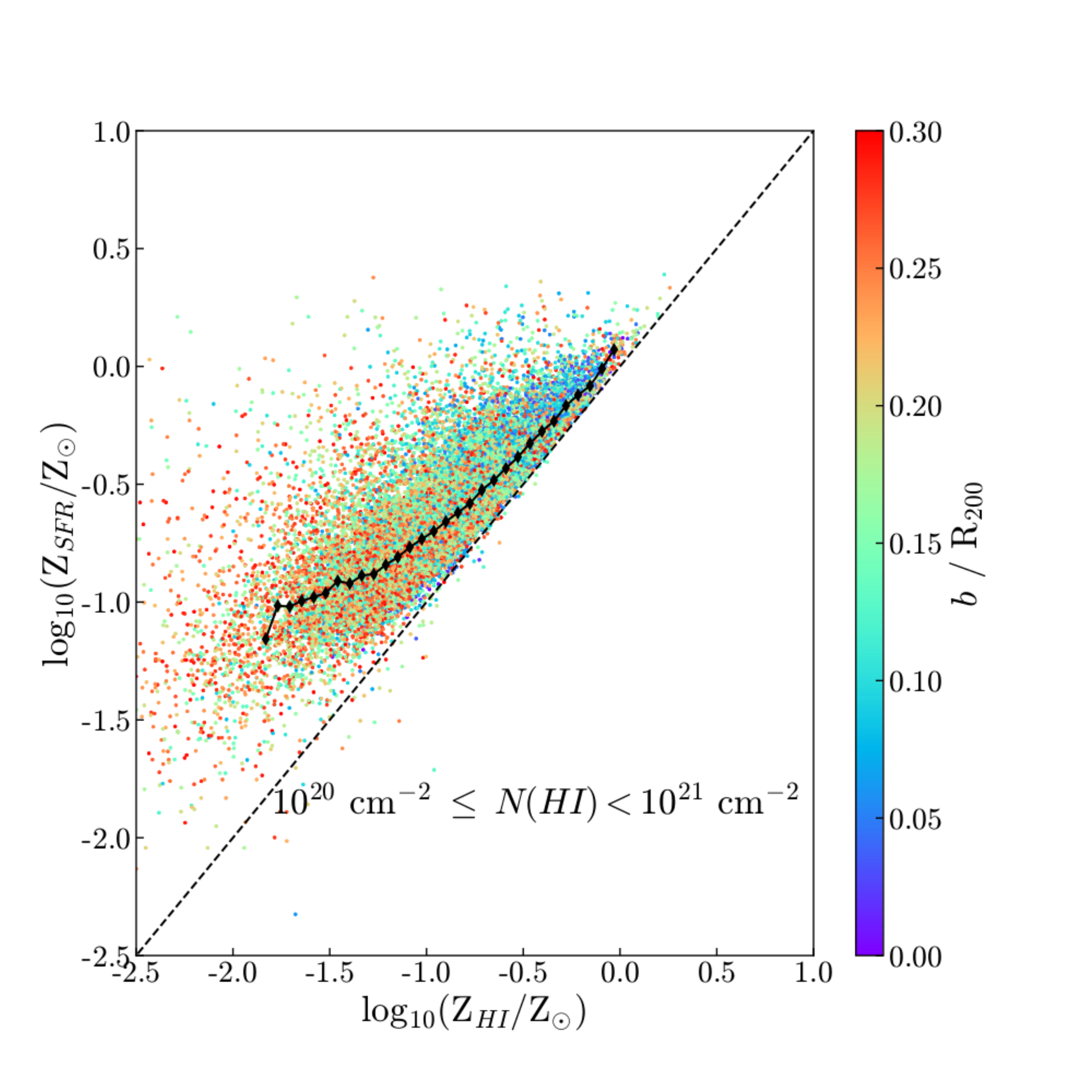} &
\includegraphics[width=0.45 \textwidth]{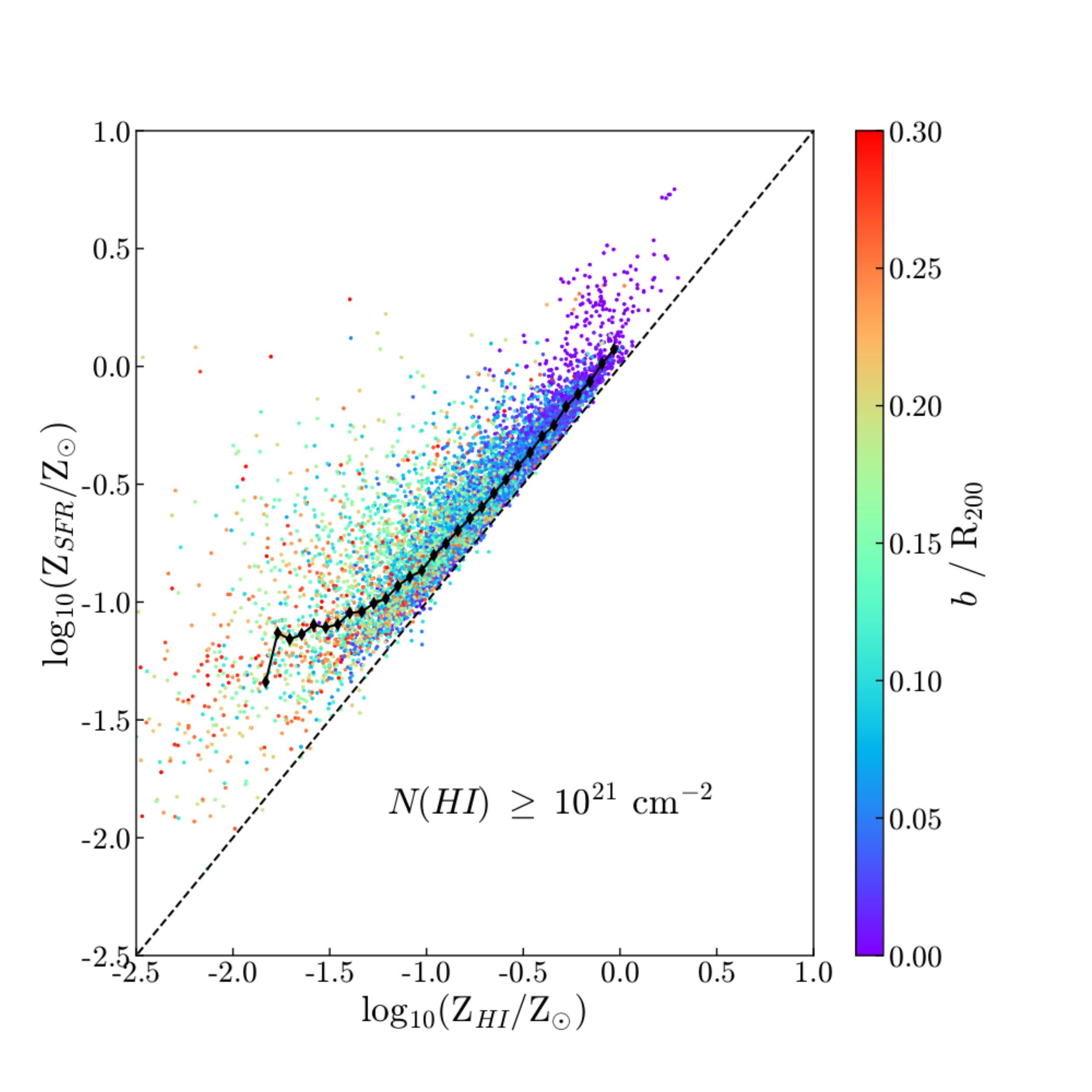} \cr
\includegraphics[width=0.45 \textwidth]{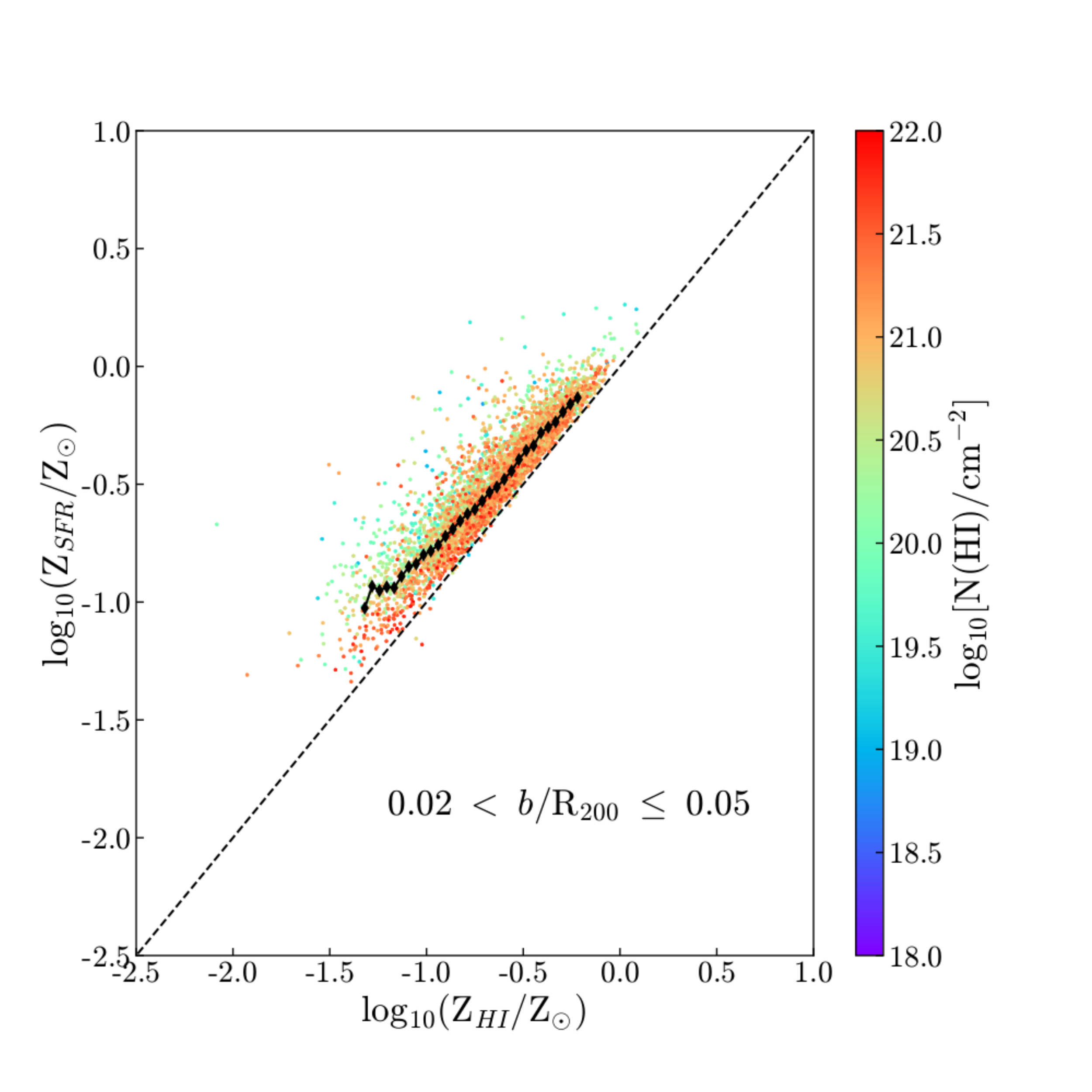} &
\includegraphics[width=0.45 \textwidth]{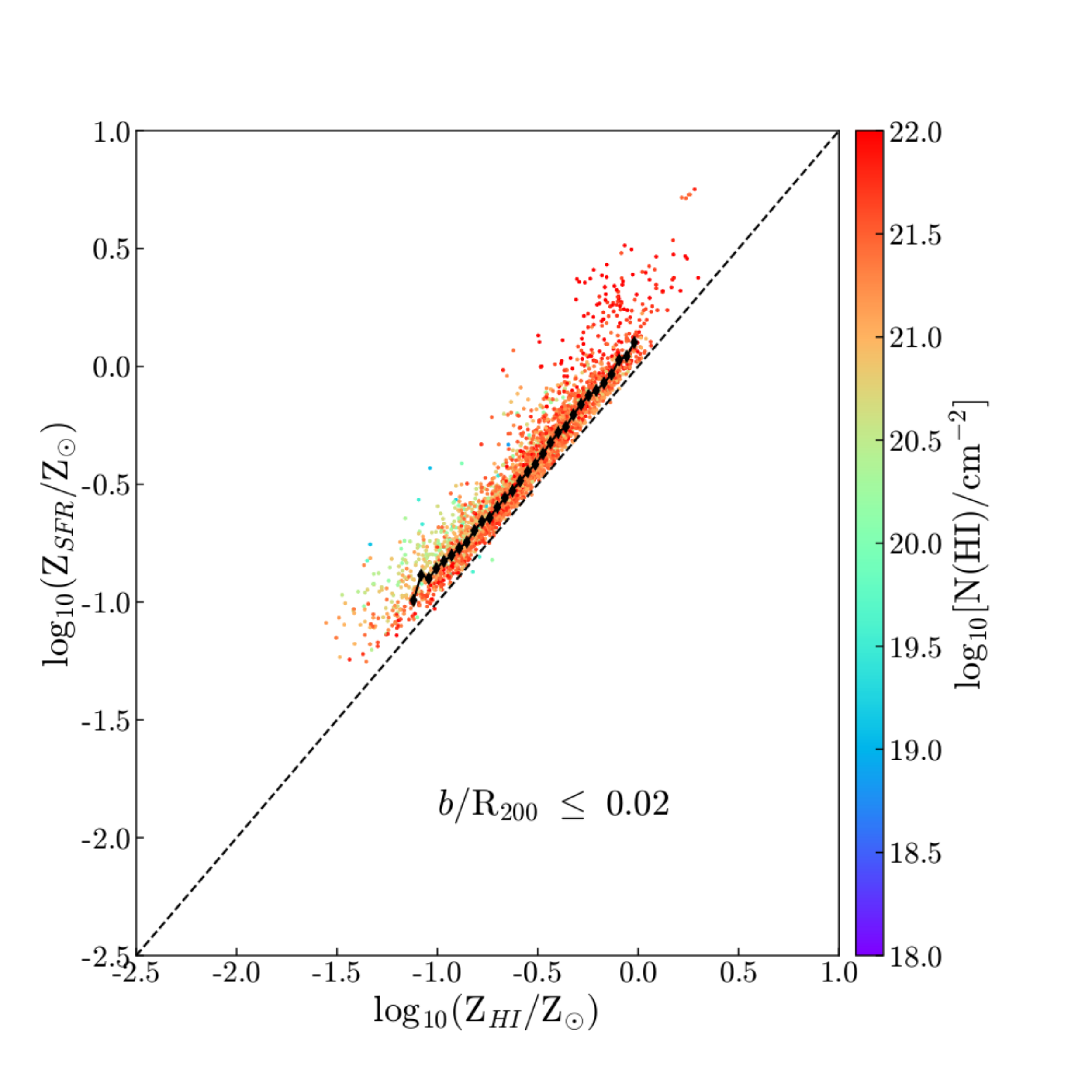} 
\end{tabular}
\caption{ 
The \zh--\zs\ plane with smoothed metallicities for  sightlines at $z=2$ selected based on N(\hi) and impact parameter. The \zsm\ values are shown with black diamonds and the x=y line is presented with the dashed lines.  
{\bf Top panels:} Sightlines with \hi\ column densities of $\rm 10^{20}\ cm^{-2} \leq $ N(\hi) $\rm < 10^{21}\ cm^{-2}$ and N(\hi) $\rm\geq 10^{21}\ cm^{-2}$ in the left and right panels respectively. All the presented sightlines are within the 0.3 R$_{200}$ from galaxy centres. The colour bar  represents the value for  $b/$R$_{200}$.  
{\bf Bottom panels:} Sightlines with $0.02 < b/$R$_{200} \leq 0.05$  in left and those within the 0.02 R$_{200}$ central regions of galaxies  in the right panel respectively. The colour bar represents the \hi\ column density of the sightlines.    
\label{fig:4panels-smoothed}} 
\end{figure*}

\begin{figure*}
\centering
\begin{tabular}{cc}
\includegraphics[width=0.45 \textwidth]{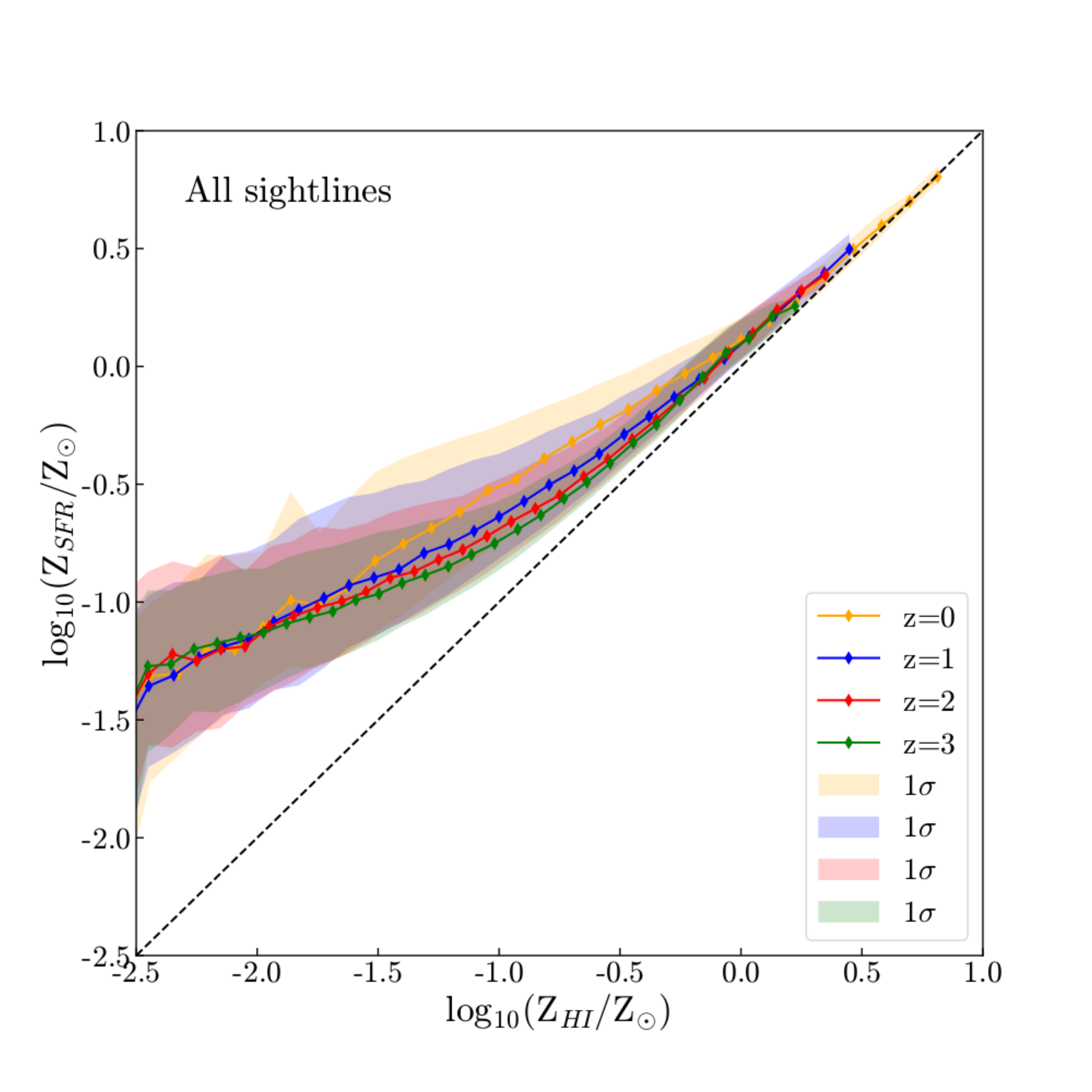} &
\includegraphics[width=0.45 \textwidth]{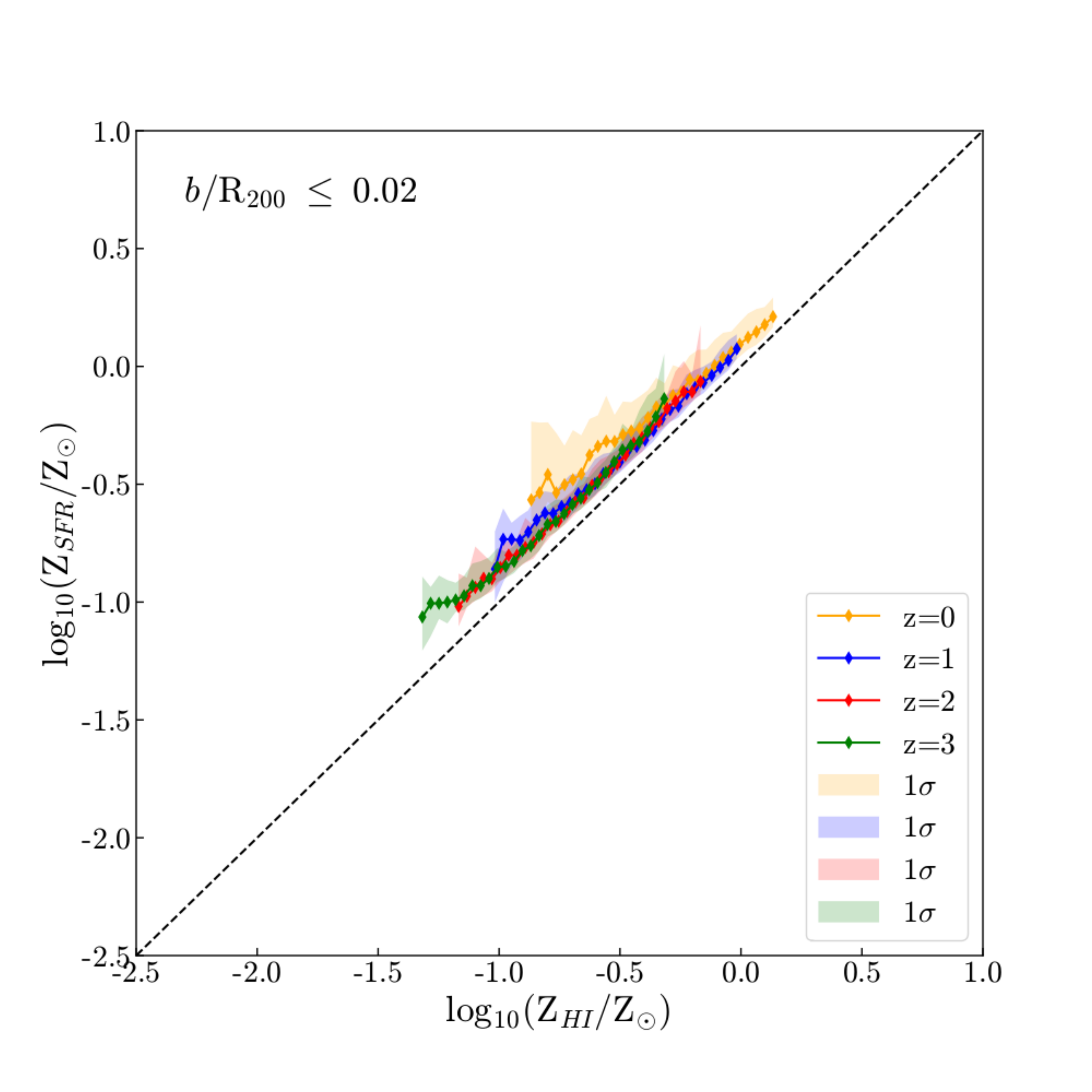} 
\end{tabular}
\caption{ 
{\bf Left:} The \zh-\zs\  relation with smoothed metallicities for all the sightlines of EAGLE galaxies at redshifts $z=0,1,2$ and 3. The \zsm\ values are marked with diamonds and the $1\sigma$ scatter around the medians are shown with shaded areas. 
{\bf Right:} The \zsm\ values of smoothed metallicities and the $1\sigma$ scatter around them for sightlines within the 0.02 R$_{200}$ central regions of galaxies in {four} redshifts.  
\label{fig:zevolution-smoothed}}
\end{figure*}

\begin{figure}
\centering
\includegraphics[width=0.5 \textwidth]{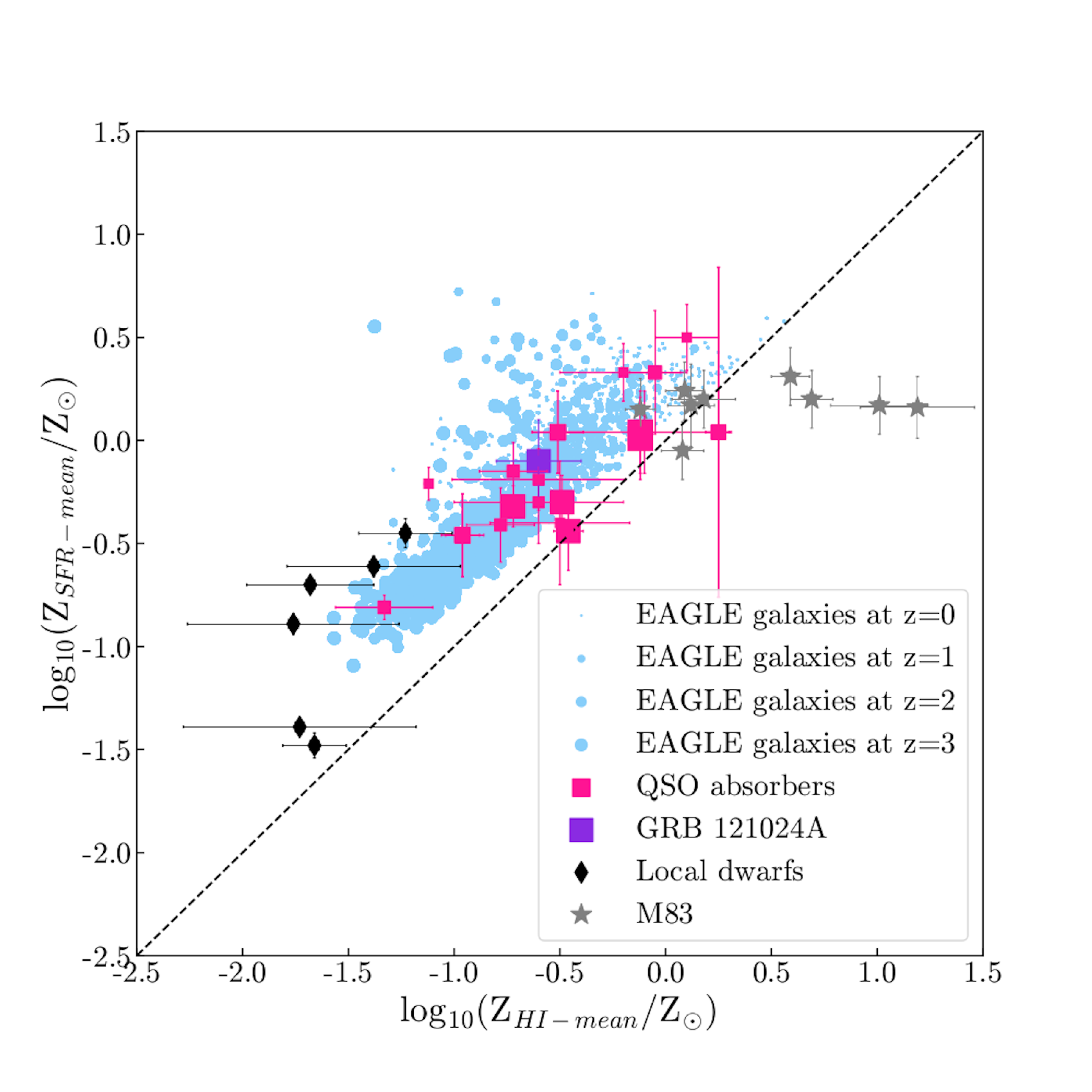}
\caption{ 
The ionised and atomic  gas metallicity measurements from observations, in comparison with the obtained metallicities of EAGLE galaxies using smoothed metallicities. 
The quasar absorbers are shown with pink squares, and  the host galaxy of GRB 121024 is presented with a purple square. 
Nearby dwarf galaxies are marked by black diamonds. The individual regions of M83 are shown by gray stars. 
The x values for these systems are measured in  single sightline  from the ratio between the column densities of heavy elements and \hi, obtained from ISM absorption lines.  The y values for these systems are obtained from the flux measurement of the bright nebular lines coming from the entire star-forming disk of the galaxy. The size of the squares are proportional to the redshift of the system. 
The blue points show the EAGLE galaxies at $z=0-3$ in the $\rm Z_{HI,mean}$--$\rm Z_{SFR,mean}$ plane.  
The quasar absorbers are shown with pink squares and the host galaxy of GRB 121024 is presented with a purple square. The x values for these systems are measured in  single sightline  from the ratio between the column densities of heavy elements and \hi, obtained from ISM absorption lines.  The y values for these systems are obtained from the flux measurement of the bright nebular lines coming from the entire star-forming disk of the galaxy. The size of the squares are proportional to the redshift of the system. 
The blue points show the EAGLE galaxies at $z=0-3$ in the $\rm Z_{HI,mean}$--$\rm Z_{SFR,mean}$ plane. 
\label{fig:obs-smooth}}
\end{figure}

\end{document}